\pdfoutput=1
\documentclass[bibyear]{aa} 

\usepackage{float}

\usepackage{graphicx}
\usepackage{txfonts}
\begin{document} 

\title{(Sub)millimetre interferometric imaging of a sample of COSMOS/AzTEC submillimetre galaxies -- 
II. The spatial extent of the radio-emitting regions}




   \author{O.~Miettinen\inst{1}, M.~Novak\inst{1}, V.~Smol\v{c}i\'{c}\inst{1}, E.~Schinnerer\inst{2}, M.~Sargent\inst{3}, E.~J.~Murphy\inst{4}, M.~Aravena\inst{5}, M.~Bondi\inst{6}, C.~L.~Carilli\inst{7,8}, A.~Karim\inst{9}, M.~Salvato\inst{10}, and G.~Zamorani\inst{11}}

   \institute{Department of Physics, University of Zagreb, Bijeni\v{c}ka cesta 32, HR-10000 Zagreb, Croatia \\ \email{oskari@phy.hr} \and 
Max-Planck-Institut f\"{u}r Astronomie, K\"{o}nigstuhl 17, 69117 Heidelberg, Germany \and Astronomy Centre, Department of Physics and Astronomy, University of Sussex, Brighton, BN1 9QH, UK \and Infrared Processing \& Analysis Center, MS 100-22, California Institute of Technology, Pasadena, CA 91125, USA \and N\'ucleo de Astronom\'{\i}a, Facultad de Ingenier\'{\i}a, Universidad Diego Portales, Av. Ej\'ercito 441, Santiago, Chile \and Istituto di Radioastronomia -- INAF, Via Gobetti 101, I-40129, Bologna, Italy \and National Radio Astronomy Observatory, P.O. Box 0, Socorro, NM 87801-0387, USA \and Astrophysics Group, Cavendish Laboratory, JJ Thomson Avenue, Cambridge CB3 0HE, UK \and Argelander-Institut f\"{u}r Astronomie, Universit\"{a}t Bonn, Auf dem H\"{u}gel 71, D-53121 Bonn, Germany \and Max-Planck-Institut f\"{u}r extraterrestrische Physik, Giessenbachstrasse 1, D-85748 Garching bei M\"{u}nchen, Germany \and INAF-Osservatorio Astronomico di Bologna, Via Ranzani 1, 40127, Bologna, Italy}

   \date{Received ; accepted}

\authorrunning{Miettinen et al.}
\titlerunning{Spatial extent of radio emission in the COSMOS SMGs}

  \abstract{Radio emission at centimetre wavelengths from highly star-forming galaxies, such as 
submillimetre galaxies (SMGs), is dominated by synchrotron radiation arising from supernova activity. 
Hence, radio continuum imaging has the potential to determine the spatial extent of star formation in these types of galaxies. 
Using deep, high-resolution ($1\sigma=2.3$~$\mu$Jy~beam$^{-1}$; $0\farcs75$) centimetre radio-continuum observations taken by 
the Karl G.~Jansky Very Large Array (VLA)-COSMOS 3~GHz Large Project, we studied the radio-emitting sizes of a flux-limited sample 
of SMGs in the COSMOS field. The target SMGs were originally discovered in a 1.1~mm continuum survey carried out with the AzTEC bolometer, 
and followed up with higher-resolution interferometric (sub)millimetre continuum observations. 
Of the 39 SMGs studied here, 3~GHz emission was detected towards 18 of them ($\sim46\pm11\%$) with signal-to-noise ratios in the range of ${\rm S/N=4.2-37.4}$. 
Towards four SMGs (AzTEC2, 5, 8, and 11), we detected two separate 3~GHz sources with projected separations of $\sim1\farcs5-6\farcs6$, but only in one or two cases (AzTEC2 and 11) they might be physically related. Using two-dimensional elliptical Gaussian fits, we derived a median deconvolved major axis FWHM size of $0\farcs54\pm 0\farcs11$ for our 18 SMGs detected at 3~GHz. For the 15 SMGs with known redshift we derived a median linear major axis FWHM of $4.2\pm0.9$~kpc. 
No clear correlation was found between the radio-emitting size and the 3~GHz or submm flux density, or the redshift of the SMG. However, there is a hint of larger radio sizes at $z\sim2.5-5$ compared to lower redshifts. The sizes we derived are consistent with previous SMG sizes measured at 1.4~GHz and in mid-$J$ CO emission, but significantly larger than those seen in the (sub)mm continuum emission (typically probing the rest-frame far-infrared with median FWHM sizes of only $ \sim1.5-2.5$~kpc). One possible scenario is that SMGs have \textit{i)} an extended gas component with a low dust temperature, and which can be traced by low- to mid-$J$ CO line emission and radio continuum emission, and \textit{ii)} a warmer, compact starburst region giving rise to the high-excitation line emission of CO, which could dominate the dust continuum size measurements. Because of the rapid cooling of cosmic-ray electrons in dense starburst galaxies ($\sim10^4-10^5$~yr), the more extended synchrotron radio-emitting size being a result of cosmic-ray diffusion seems unlikely. Instead, if SMGs are driven by galaxy mergers -- a process where the galactic magnetic fields can be pulled out to larger spatial scales -- the radio synchrotron emission might arise from more extended magnetised interstellar medium around the starburst region.} 


   \maketitle
%

\section{Introduction}

Submillimetre galaxies (SMGs; e.g. \cite{smail1997}; \cite{hughes1998}; \cite{barger1998}) 
represent a population of distant galaxies where star formation is heavily obscured by 
the dusty interstellar medium (ISM). The star formation rates (SFRs) in SMGs lie in the range 
of $\sim10^2-10^3$~M$_{\sun}$~yr$^{-1}$, and hence these galaxies stand out as the most intense 
starbursts in the universe (for reviews, see \cite{blain2002}; \cite{casey2014}). 
 As the potential precursors to the present-day massive elliptical galaxies, SMGs have become one of the primary targets for 
understanding galaxy evolution across cosmic time (e.g. \cite{swinbank2006}; \cite{fu2013}; \cite{toft2014}; \cite{simpson2014}). In the context of this evolutionary connection, 
determining the sizes and size evolution of SMGs is crucial.

Nearly all of the cm-wavelength radio emission from star-forming galaxies, such as SMGs, is non-thermal synchrotron radiation from relativistic electrons accelerated in supernova (SN) remnants produced by the short-lived, high-mass OB-type stars ($M\gtrsim8$~M$_{\sun}$; main-sequence lifetime $\tau_{\rm MS}\lesssim30$~Myr). 
Because SNe are tracing the recent/on-going star formation, the radio synchrotron emission has the potential to trace the spatial scales on which star formation is occurring. 
This connection between radio emission and star formation is strongly supported by 
 the tight infrared (IR)-radio correlation observed in galaxies (e.g. \cite{helou1985}; \cite{beck1988}; \cite{xu1992}; 
\cite{condon1992}; \cite{yun2001}; \cite{bell2003}; \cite{tabatabaei2007}; \cite{murphy2008}; \cite{sargent2010}; \cite{moric2010}; \cite{dumas2011}). 
On the basis of this correlation, the IR-emitting region of a star-forming galaxy is expected to be comparable in size to that of radio continuum emission. However, 
the most recent studies of the sizes of IR-emitting regions of SMGs based on continuum imaging observations with the Atacama Large Millimetre/submillimetre Array (ALMA) show 
that these are significantly smaller than SMG radio sizes presented in the literature (\cite{simpson2015a}; \cite{ikarashi2015}). A possible explanation for this discrepancy, as suggested 
by Simpson et al. (2015a), is cosmic ray (CR) diffusion in the galactic magnetic field away from their acceleration site, which would render larger radio sizes. To test this further here 
we present a study of radio sizes of SMGs from a well selected sample of SMGs in the Cosmic Evolution Survey (COSMOS; \cite{scoville2007}) deep field using radio data from the 
Karl G.~Jansky Very Large Array (VLA)-COSMOS 3~GHz Large Project ($1\sigma$ noise of 2.3~$\mu$Jy~beam$^{-1}$, angular resolution $0\farcs75$; V.~Smol{\v c}i{\'c} et al., in prep.). We describe 
the SMG sample and the employed VLA data in detail in Sect.~2. The 3~GHz images are presented in Sect.~3, and the analysis (size measurements and radio spectral indices) are presented in Sect.~4. 
We compare our results with literature studies in Sect.~5, discuss the results in Sect.~6, and summarise the main results of the paper in Sect.~7.

To be consistent with the most recent results from the \textit{Planck} mission (\cite{planck2015}), 
the cosmology adopted in the present work corresponds to the flat $\Lambda$CDM universe with 
the dark energy density $\Omega_{\Lambda}=0.692$, total (dark+luminous baryonic) matter density 
$\Omega_{\rm m}=0.308$, and a Hubble constant of $H_0=67.8$ km~s$^{-1}$~Mpc$^{-1}$.


\section{Data}

\subsection{Source sample}

The target SMGs of the present study -- AzTEC1--30 -- were originally discovered in the JCMT/AzTEC 1.1 mm continuum survey 
($18\arcsec$ resolution) towards a COSMOS subfield (0.15~deg$^2$ in size) by Scott et al. (2008). 
The signal-to-noise ratios of these SMGs were found to be in the range of S/N$_{\rm 1.1\, mm}=4.0-8.3$ 
(see Table~1 in \cite{scott2008}). The 15 brightest sources, AzTEC1--15 (S/N$_{\rm 1.1\, mm} \geq 4.6$), 
were imaged (and detected) with the Submillimetre Array (SMA) at 890 $\mu$m ($2\arcsec$ resolution) by 
Younger et al. (2007, 2009). More recently, AzTEC16--30 (S/N$_{\rm 1.1\, mm}=4.0-4.5$) were 
imaged with the Plateau de Bure Interferometer (PdBI) at 1.3~mm ($\sim 1\farcs8$ resolution) by Miettinen et al. (2015). 
These interferometric follow-up studies have allowed us to accurately determine the position of the actual SMGs giving rise 
to the millimetre continuum emission seen in the single-dish AzTEC maps and, in eight cases, to resolve the single-dish emission into 
multiple (two to three) components (at $\sim2\arcsec$ resolution). 
This way, we can reliably identify the correct 3~GHz counterparts of the target SMGs. We note that even the faintest component 
in our source sample (AzTEC26b) has a 1.3~mm flux density of 0.9~mJy, which corresponds to $\sim4$~mJy at the observed-frame 850~$\mu$m (assuming a dust emissivity index of $\beta=1.5$; see 
\cite{miettinen2015}), and hence can be considered an SMG [cf.~the classic SMG threshold of $S_{\rm 850\, \mu m}>5$~mJy refers to bright SMGs (e.g. \cite{hainline2009}; \cite{gonzalez2011})]. 
We also note that none of these SMGs has been detected in X-rays, and hence they do not appear to harbour any strong active galactic nucleus (AGN)
[a typical $3\sigma$ upper limit to the flux density in the 0.5--2~keV band data of the \textit{Chandra} COSMOS Legacy Survey is 
$<6 \times 10^{-16}$~erg~cm$^{-2}$~s$^{-1}$ (F.~Civano et al., in prep.)]. 
This suggests that the observed radio emission from our SMGs is predominantly powered by star formation. 
This is further supported by the fact that none of our SMGs were detected with the Very Long Baseline Array (VLBA) observations 
at a high, milliarcsec resolution at 1.4~GHz (N.~Herrera Ruiz et al., in prep.), yielding a $3\sigma$ flux density upper limit to 
$S_{\rm 1.4\, GHz}^{\rm VLBA}$ of $<60$~$\mu$Jy~beam$^{-1}$.\footnote{As described in Appendix~A, we have detected two 3~GHz sources towards AzTEC8. 
The western radio source is associated with our target SMG, while the eastern 3~GHz source, physically unrelated to the SMG, is also detected at 1.4~GHz with the VLBA.}

\begin{figure}[!h]
\centering
\resizebox{\hsize}{!}{\includegraphics{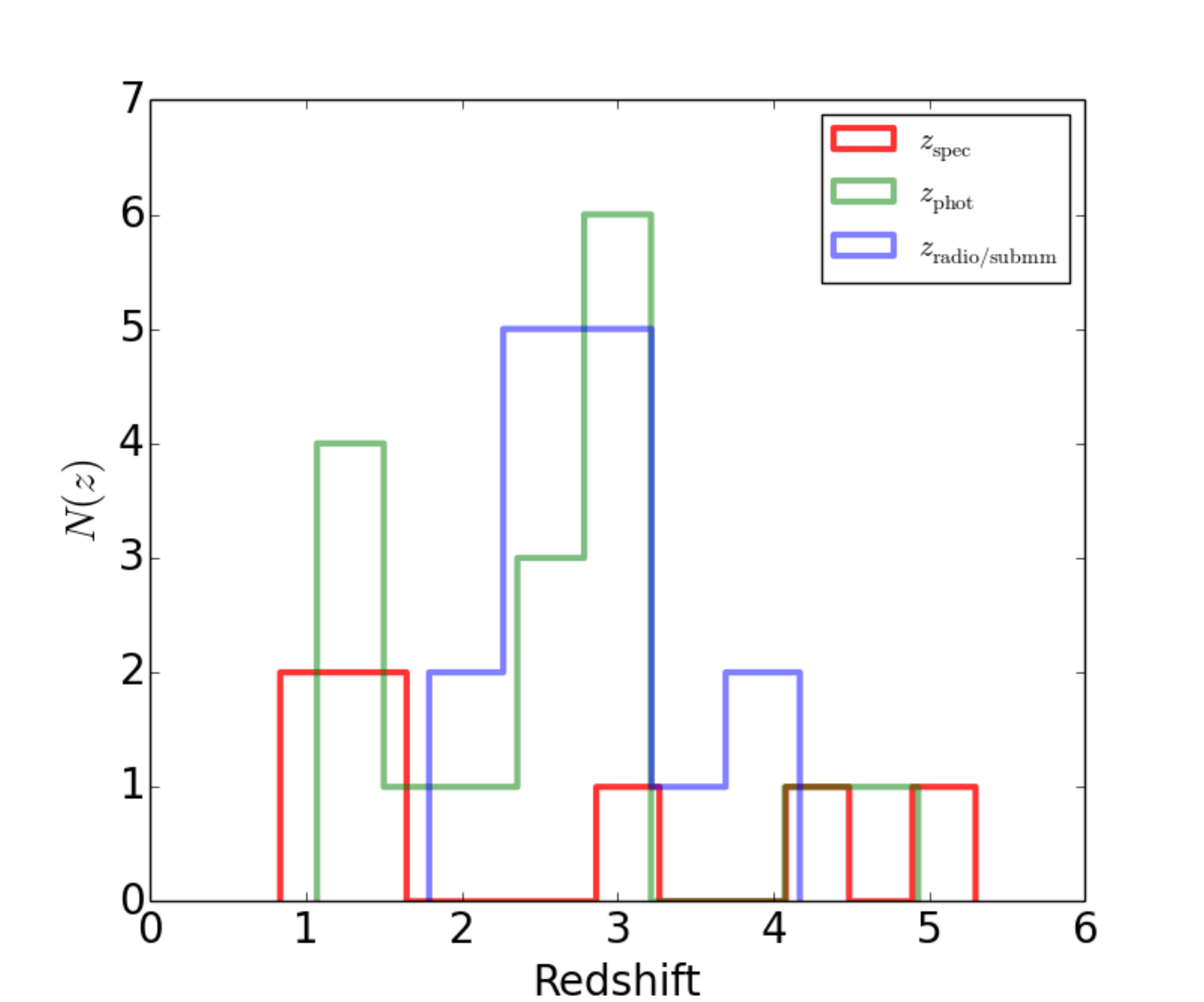}}
\caption{Redshift distribution of the target SMGs divided into three subsamples: the red and green histograms show the spectroscopic and photometric redshifts, 
respectively, while the blue histogram shows the redshift values derived from the radio-to-submm spectral index (see Table~\ref{table:sample}). 
The redshift bins have a width of $\Delta z=0.4$. The lower redshift limits were placed in the bins corresponding to those values.}
\label{figure:zdist}
\end{figure}

Our sample of 39 SMGs is listed in Table~\ref{table:sample}. The coordinates given in the table correspond to the (sub)mm peak positions 
determined in the aforementioned SMA and PdBI studies. Table~\ref{table:sample} also provides the source redshifts that are 
based on spectroscopic measurements (seven sources), optical to near-infrared (NIR) spectral energy distribution fitting 
(i.e. photometric redshift; 17 sources), and radio/submm flux density ratios (15 sources). 
The redshift distribution is shown in Fig.~\ref{figure:zdist}. We refer to Miettinen et al. (2015 and references therein) for further 
details and discussion on the redshifts of our SMGs. 

\begin{table}
\renewcommand{\footnoterule}{}
\caption{Source list.}
{\scriptsize
\begin{minipage}{1\columnwidth}
\centering
\label{table:sample}
\begin{tabular}{c c c c c}
\hline\hline 
Source ID & $\alpha_{2000.0}$ & $\delta_{2000.0}$ & Redshift\tablefootmark{a} & $z$ reference\tablefootmark{a}\\
       & [h:m:s] & [$\degr$:$\arcmin$:$\arcsec$] & & \\ 
\hline
AzTEC1 &  09 59 42.86 & +02 29 38.2 & $z_{\rm spec}=4.3415$ & 1 \\
AzTEC2 &  10 00 08.05 & +02 26 12.2 & $z_{\rm spec}=1.125$ & 2 \\
AzTEC3 &  10 00 20.70 & +02 35 20.5 & $z_{\rm spec}=5.298$ & 3\\
AzTEC4 &  09 59 31.72 & +02 30 44.0 & $z_{\rm phot}=4.93_{-1.11}^{+0.43}$ & 4 \\
AzTEC5 &  10 00 19.75 & +02 32 04.4 & $z_{\rm phot}=3.05_{-0.28}^{+0.33}$ & 4 \\
AzTEC6 &  10 00 06.50 & +02 38 37.7 & $z_{\rm radio/submm}>3.52$ & 5 \\
AzTEC7 &  10 00 18.06 & +02 48 30.5 & $z_{\rm phot}=2.30\pm0.10$ & 4 \\
AzTEC8 &  09 59 59.34 & +02 34 41.0 & $z_{\rm spec}=3.179$ & 6  \\
AzTEC9 &  09 59 57.25 & +02 27 30.6 & $z_{\rm phot}=1.07_{-0.10}^{+0.11}$ & 4 \\
AzTEC10 &  09 59 30.76 & +02 40 33.9 & $z_{\rm phot}=2.79_{-1.29}^{+1.86}$ & 4 \\
AzTEC11-N\tablefootmark{b} & 10 00 08.91 & +02 40 09.6 & $z_{\rm spec}=1.599$ & 7  \\
AzTEC11-S\tablefootmark{b} & 10 00 08.94 & +02 40 12.3 & $z_{\rm spec}=1.599$ & 7 \\
AzTEC12 &  10 00 35.29 & +02 43 53.4 & $z_{\rm phot}=2.54_{-0.33}^{+0.13}$ & 4 \\
AzTEC13 &  09 59 37.05 & +02 33 20.0 & $z_{\rm radio/submm}>4.07$ & 5 \\
AzTEC14-E\tablefootmark{c} & 10 00 10.03 & +02 30 14.7 & $z_{\rm radio/submm}>2.95$ & 5\\
AzTEC14-W\tablefootmark{c} & 10 00 09.63 & +02 30 18.0 & $z_{\rm phot}=1.30_{-0.36}^{+0.12}$ & 4 \\
AzTEC15 & 10 00 12.89 & +02 34 35.7 & $z_{\rm phot}=3.17_{-0.37}^{+0.29}$ & 4  \\
AzTEC16 & 09 59 50.069 & +02 44 24.50 & $z_{\rm radio/submm}>2.42$ & 5 \\
AzTEC17a & 09 59 39.194 & +02 34 03.83 & $z_{\rm spec}=0.834$ & 7 \\
AzTEC17b & 09 59 38.904 & +02 34 04.69 & $z_{\rm phot}=4.14_{-1.73}^{+0.87}$ & 5 \\
AzTEC18 & 09 59 42.607 & +02 35 36.96 & $z_{\rm phot}=3.00_{-0.17}^{+0.19}$ & 5 \\
AzTEC19a & 10 00 28.735 & +02 32 03.84 & $z_{\rm phot}=3.20_{-0.45}^{+0.18}$ & 5 \\
AzTEC19b & 10 00 29.256 & +02 32 09.82 & $z_{\rm phot}=1.11\pm0.10$ & 5 \\ 
AzTEC20 & 10 00 20.251 & +02 41 21.66 & $z_{\rm radio/submm}>2.35$ & 5  \\
AzTEC21a & 10 00 02.558 & +02 46 41.74 & $z_{\rm phot}=2.60_{-0.17}^{+0.18}$ & 5 \\
AzTEC21b & 10 00 02.710 & +02 46 44.51 & $z_{\rm phot}=2.80_{-0.16}^{+0.14}$ & 5 \\
AzTEC21c & 10 00 02.856 & +02 46 40.80 & $z_{\rm radio/submm}>1.93$ & 5  \\
AzTEC22 & 09 59 50.681 & +02 28 19.06 & $z_{\rm radio/submm}>3.00$ & 5  \\ 
AzTEC23 & 09 59 31.399 & +02 36 04.61 & $z_{\rm phot}=1.60_{-0.50}^{+0.28}$ & 5 \\
AzTEC24a & 10 00 38.969 & +02 38 33.90 & $z_{\rm radio/submm}>2.35$ & 5  \\
AzTEC24b & 10 00 39.410 & +02 38 46.97 & $z_{\rm radio/submm}>2.28$ & 5  \\
AzTEC24c & 10 00 39.194 & +02 38 54.46 & $z_{\rm radio/submm}>3.17$ & 5  \\
AzTEC25\tablefootmark{d} & \ldots & \ldots & \ldots & \ldots\\
AzTEC26a & 09 59 59.386 & +02 38 15.36 & $z_{\rm phot}=2.50_{-0.14}^{+0.24}$ & 5 \\
AzTEC26b & 09 59 59.657 & +02 38 21.08 & $z_{\rm radio/submm}>1.79$ & 5  \\
AzTEC27 & 10 00 39.211 & +02 40 52.18 & $z_{\rm radio/submm}>4.17$ & 5 \\
AzTEC28 & 10 00 04.680 & +02 30 37.30 & $z_{\rm radio/submm}>3.11$ & 5  \\
AzTEC29a & 10 00 26.351 & +02 37 44.15 & $z_{\rm radio/submm}>2.96$ & 5 \\
AzTEC29b & 10 00 26.561 & +02 38 05.14 & $z_{\rm phot}=1.45_{-0.38}^{+0.79}$ & 5 \\
AzTEC30 & 10 00 03.552 & +02 33 00.94 & $z_{\rm radio/submm}>2.51$ & 5 \\
\hline 
\end{tabular} 
\tablefoot{The coordinates given in columns~(2) and (3) for AzTEC1--15 refer to the SMA 890 $\mu$m peak position (\cite{younger2007}, 2009), 
while those for AzTEC16--30 are the PdBI 1.3 mm peak positions (\cite{miettinen2015}). 
\tablefoottext{a}{The $z_{\rm spec}$, $z_{\rm phot}$, and $z_{\rm radio/submm}$ values are the spectroscopic redshift, optical-NIR photometric redshift, and the redshift derived using the Carilli-Yun redshift indicator (\cite{carilli1999}, 2000). The $z$ references in the last column are as follows: $1=$\cite{yun2015}; $2=$M.~Balokovi\'c et al., in prep.; $3=$\cite{riechers2010} and \cite{capak2011}; $4=$\cite{smolcic2012}; $5=$\cite{miettinen2015}; $6=$D.~A.~Riechers et al., in prep.; $7=$M.~Salvato et al., in prep.}\tablefoottext{b}{AzTEC11 was resolved into two 890 $\mu$m sources (N and S) by Younger et al. (2009). The two components are probably physically related, i.e. are at the same redshift (see Appendix~A).}\tablefoottext{c}{AzTEC14 was resolved into two 890 $\mu$m sources (E and W) by Younger et al. (2009). The eastern component appears to lie at a higher redshift than the western one (\cite{smolcic2012}).}\tablefoottext{d}{AzTEC25 was not detected in the 1.3~mm PdBI observations (\cite{miettinen2015}).}
}
\end{minipage} 
}
\end{table}

\subsection{VLA 3~GHz radio continuum data}

The observations used in the present paper were taken by the VLA-COSMOS 3~GHz Large Project (PI: V.~Smol{\v c}i{\'c}; 
V.~Smol{\v c}i{\'c} et al., in prep.). Details of the observations, data reduction, and imaging can be found in Novak et al. (2015), 
Smol{\v c}i{\'c} et al. (2015a), and V.~Smol{\v c}i{\'c} et al. (in prep.). In the present paper, we employ -- for the first time -- 
the final, full 3~GHz mosaic imaging of COSMOS (192 pointings in total). Briefly, these S-band observations were carried out with the VLA of
the NRAO\footnote{The National Radio Astronomy Observatory is a facility of the National Science Foundation operated under cooperative 
agreement by Associated Universities, Inc.} in its A and C configurations (maximum baseline of 36.4~km and 3.4~km, respectively) 
between 2012 and 2014. The 2~GHz bandwidth (2 basebands of 1~GHz each) used was divided into 16 sub-bands/spectral windows (SPWs) 
each with a 128~MHz bandwidth. Each SPW was subdivided into 64 spectral channels with a width of 2~MHz. The data were calibrated using the 
AIPSLite data reduction pipeline, which is an extension of NRAO's Astronomical Image Processing System (AIPS)\footnote{{\tt http://www.aips.nrao.edu/index.shtml}} package
(\cite{bourke2014}; K.~Mooley et al., in prep.), and adapted for the VLA-COSMOS 3~GHz Large Project (for details, see V.~Smol{\v c}i{\'c} et al., in prep.).
Further editing, flagging, and imaging was done using the Common Astronomy Software Applications package (CASA\footnote{CASA is developed by 
an international consortium of scientists based at the NRAO, the European Southern Observatory (ESO), the National Astronomical Observatory of Japan (NAOJ), 
the CSIRO Australia Telescope National Facility (CSIRO/ATNF), and the Netherlands Institute for Radio Astronomy (ASTRON) under the guidance of NRAO. 
See {\tt http://casa.nrao.edu}}; \cite{mcmullin2007}). To reduce sidelobes and artefacts in the data, phase solutions obtained from 
self-calibration with the bright quasar J1024-0052 were applied on each pointing. Every field was cleaned down to $5\sigma$, and further 
cleaned down to $1.5\sigma$ using manually defined masks around the sources.

The data used here were imaged using the multi-scale multi-frequency synthesis (MS-MFS) method 
(\cite{rau2011}). Briggs or robust weighting was applied to the calibrated visibilities with a robust value of 0.5.
Considering the aim of the present study (i.e. measuring the 3~GHz sizes of our SMGs), the main advantage of MS-MFS is that 
the final image resolution is not determined by the lowest frequency of the bandwidth used because all the SPWs are used 
in the image deconvolution. A Gaussian $u$-$v$ tapering was applied on each pointing using their own Gaussian beam size 
(Full Width at Half Maximum or FWHM). The final mosaic was restored with a circular synthesised beam size (FWHM) of 
$\theta_{\rm maj}=\theta_{\rm min}=0\farcs75$, where $\theta_{\rm maj}$ and $\theta_{\rm min}$ are the major and minor axes of the beam. 
The final $1\sigma$ root mean square (rms) noise level in our maps is typically about 2.3~$\mu$Jy~beam$^{-1}$. 

To quantify the effect of bandwidth smearing (BWS) in our 3~GHz mosaic, we examined the behaviour of the ratio of the total integrated source flux density 
to its peak surface brightness as a function of the S/N ratio (e.g. \cite{bondi2008}; \cite{novak2015}; V.~Smol{\v c}i{\'c} et al., in prep.). 
This comparison showed that the effect of BWS in the full 3~GHz mosaic of COSMOS is only up to $\sim 3\%$, and no correction for BWS in the peak 
surface brightness is applied in the present study. To further examine the importance of BWS, we created images of a subsample of our sources 
from separate pointings where the source distance from the (nearest) phase centre is different. Besides depending on the fractional bandwidth, 
the magnitude of BWS is directly proportional to the angular distance of the source from the phase centre. 
However, no significant radial smearing was seen in the aforementioned images, which lends further support to negligible BWS.


\section{3~GHz images and counterpart identification of the AzTEC SMGs} 


The 3~GHz images towards our SMGs are shown in Fig.~\ref{figure:maps}. We note that at the redshifts of our sources, $z=0.834-5.298$, 
we are probing rest-frame frequencies of $\nu_{\rm rest}\simeq5.5-19$~GHz ($\lambda_{\rm rest}\simeq1.6-5.5$~cm), which are dominated by non-thermal 
synchrotron radiation with the fraction of thermal emission becoming increasingly important at higher frequencies 
(e.g. \cite{condon1992}; \cite{murphy2012b}). The 3~GHz counterparts of our SMGs were identified by eye inspection of the corresponding images. 
The SMGs AzTEC1--9, 11-N and 11-S, 12, 15, 17a, 19a, 21a, 24b, and 27 are found to be associated with a 3~GHz source 
(with a median offset of $0\farcs26$; Table~\ref{table:results}), i.e. 18/39 or $\sim46\%$ (with a Poisson error on counting statistics of $\pm11\%$) 
of our sources are 3 GHz-emitting SMGs. The S/N ratios of our detected 3~GHz sources are in the range of ${\rm S/N=4.2-37.4}$, 
AzTEC7 being the most significant detection. We note that the detection S/N ratio at 1.1~mm of these 3~GHz-emitting SMGs was found to be in the range of 
${\rm S/N}_{\rm 1.1\, mm}=4.0-8.3$ (\cite{scott2008}). To summarise, 18 SMGs in our sample are found to be associated with 3~GHz emission. 
A selection of these SMGs, and the additional 3~GHz radio sources not analysed further in the present study are discussed 
in more detail in Appendices~A and B, respectively. The SMGs not detected at 3~GHz are discussed in Appendix~C.

\begin{figure*}[!htb]
\begin{center}
\includegraphics[width=\textwidth]{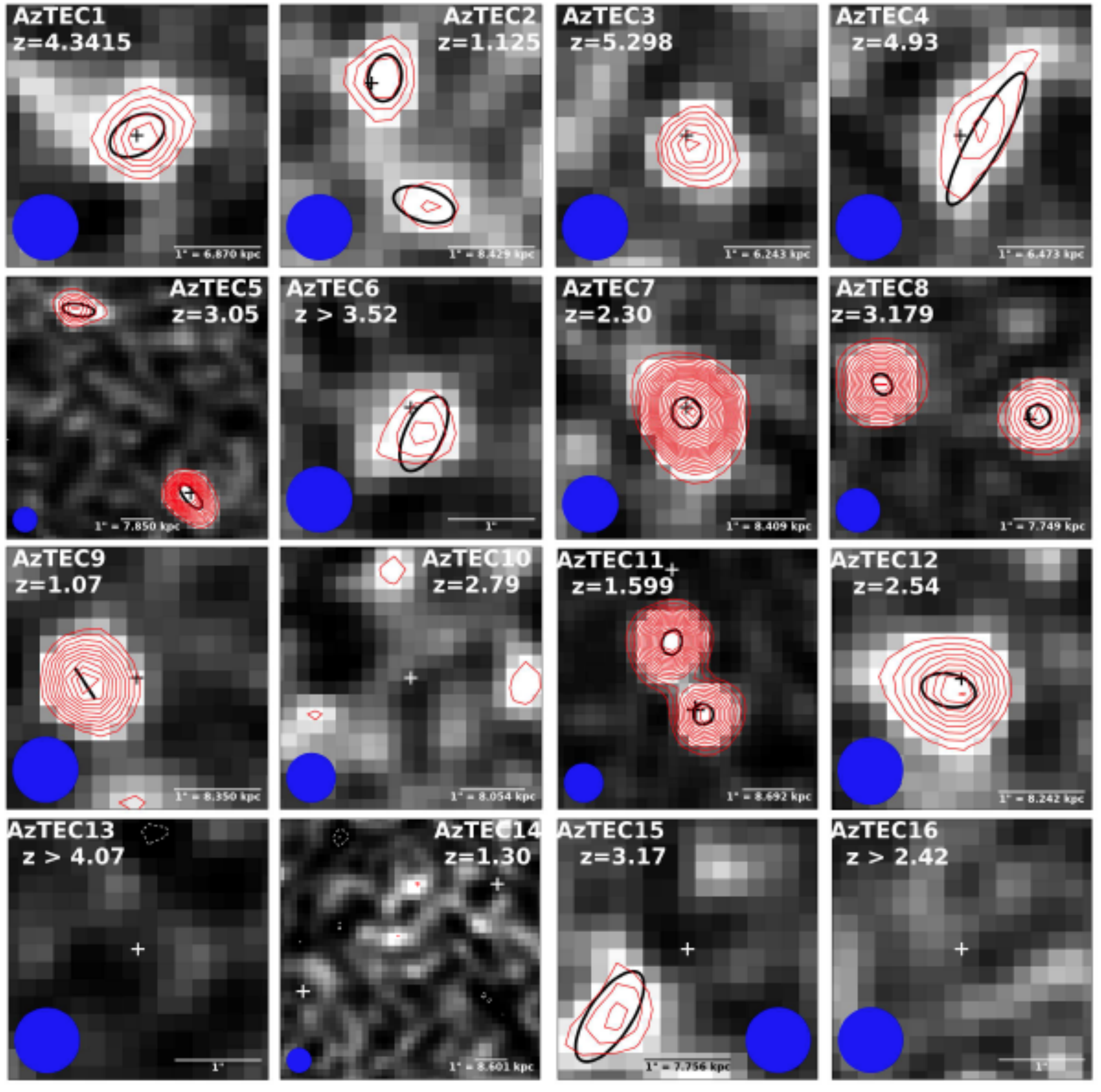}
\caption{The VLA 3~GHz images towards AzTEC1--30 displayed with north up and east left. The greyscale images are shown with power-law scaling 
(except for AzTEC28, where arcsinh scaling is used to better illustrate the intensity scale; a power-law scaling would be completely black), 
and the overlaid red contours start from $3\sigma$ and increase in steps of $1\sigma$ except for AzTEC5, 7, 8, 11, 12, 22, and 24 where the step is $\sqrt{2}\times\sigma$. 
The white dashed contours show the corresponding negative features (starting from $-3\sigma$). The plus sign in each panel marks the (sub)mm 
peak position (SMA 890~$\mu$m for AzTEC1--15: PdBI 1.3~mm for AzTEC16--30). The black thick ellipse shows the resulting Gaussian fit to the source 
(centred at the peak position, FWHM size, and P.A.). For AzTEC2 (both components), 3, 4, 5 (both components), 7, 8 (both components), 11-N, 11-S, and 12 
the size represents an upper limit (see Table~\ref{table:results}). Moreover, for AzTEC4 the peak position was not well determined, and for 
AzTEC9, 17a, and 27 only the major axis FWHM could be determined by {\tt JMFIT}. The blue filled circle shows the synthesised beam size ($0\farcs75$ FWHM). Note that the areal coverage of the images differs from each other for illustrative purposes; a scale bar indicating the $1\arcsec$ projected length is shown in each panel, annotated with the corresponding proper length [kpc] at the quoted SMG redshift (except when only a lower limit to $z$ is available).}
\label{figure:maps}
\end{center}
\end{figure*}

\addtocounter{figure}{-1} 
\begin{figure*}[!htb]
\begin{center}
\includegraphics[width=\textwidth]{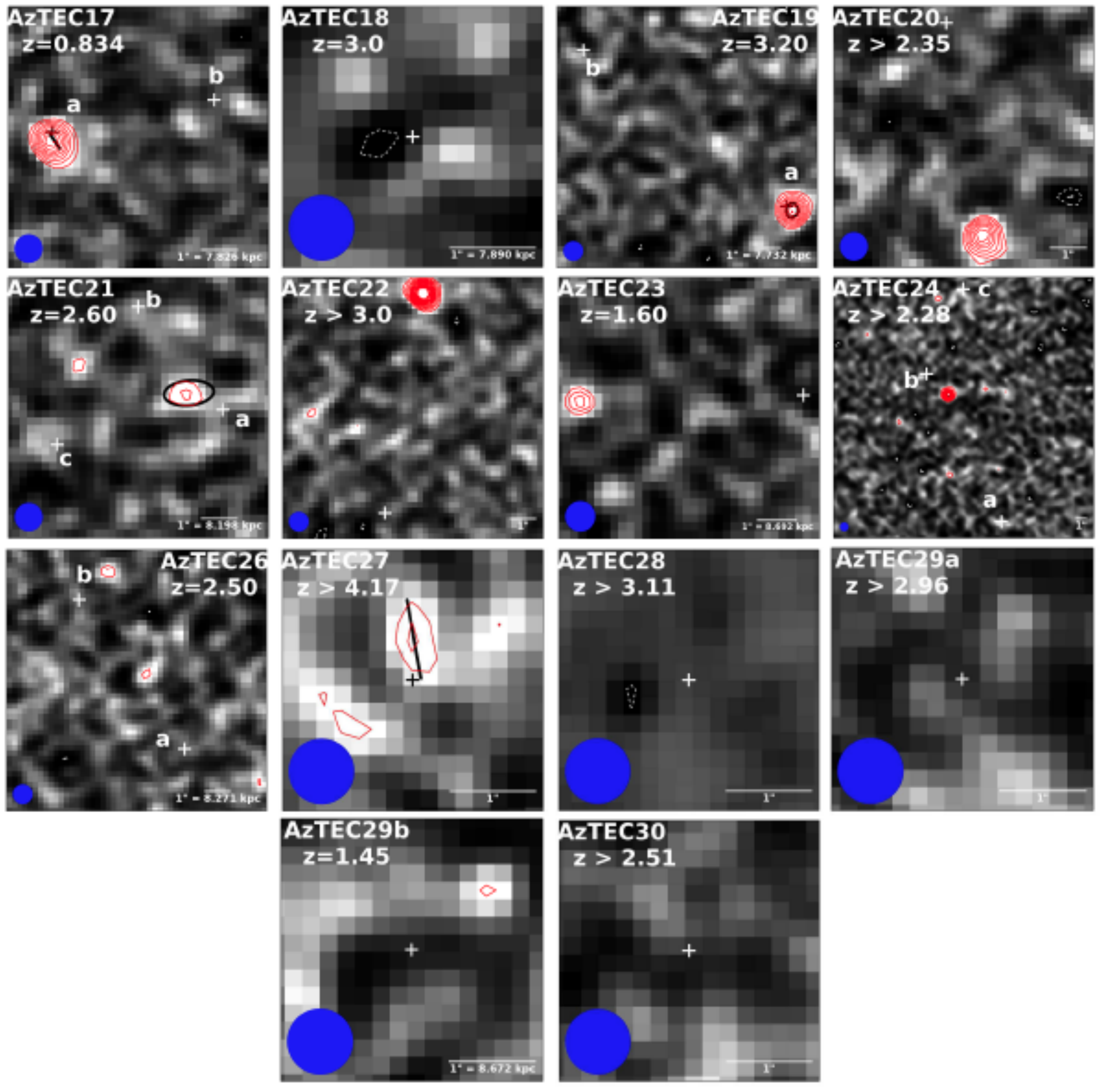}
\caption{continued.}
\label{figure:maps}
\end{center}
\end{figure*}

\section{Analysis} 

\subsection{Measuring the size of the radio-emitting region} 

We used AIPS package to determine the deconvolved sizes of our 3~GHz sources. Two-dimensional elliptical Gaussian fits to the image plane data were made 
using the AIPS task {\tt JMFIT}. The fitting was performed inside a box containing the source, and the fit was restricted 
to the pixel values of $\geq2.5\sigma$. The results are given in Table~\ref{table:results}, and illustrated in Fig.~\ref{figure:maps}. 
 To test the reliability of our size measurements, we simulated SMGs with an assigned size and varying S/N ratio, and fit them in the same manner as the real sources. 
These simulations, described in Appendix~D, suggest that the sizes provided by {\tt JMFIT} are generally robust within the uncertainties assigned by the fitting task 
(see Fig.~\ref{figure:sim}, lower panel). As is often done in radio-continuum surveys, we considered a source to be resolved if its deconvolved FWHM size is larger 
than one-half the synthesised beam FWHM (e.g. \cite{mundell2000}; \cite{ho2001}; \cite{urquhart2009}). AzTEC1, 6, 15, 19a, 21a, and 24b are resolved 
in both $\theta_{\rm maj}$ and $\theta_{\rm min}$, while AzTEC2 (both components), 4, 5 (both components), 7, 8, 11-N, 11-S, and 12 are resolved in $\theta_{\rm maj}$ but 
unresolved in $\theta_{\rm min}$. AzTEC3 and the additional component towards AzTEC8 are unresolved in both axes. 
The upper size limit for unresolved sources was set to one-half the synthesised beam FWHM ($<0\farcs38$). For AzTEC9, 17a, and 27 only 
the major axis FWHM could be determined by {\tt JMFIT}, while the fitting task did not provide a value for the minor axis FWHM (the output value$=0$). 
  In column~(8) in Table~\ref{table:results}, we give the projected linear FWHM size for those SMGs with known redshift 
[i.e. not just a lower limit to $z$ derived using the Carilli \& Yun (1999, 2000) method]. 
As mentioned earlier in Sect.~2.2, no correction for the negligible BWS in the peak surface brightness or FWHM size was applied. 

To calculate the statistical properties of our radio-emission size distribution, we applied survival analysis to take the upper limits to the size
into account. We assumed that the censored data follow the same distribution as the uncensored values, and we used 
the Kaplan-Meier (K-M) method to construct a model of the input data [for this purpose, we used the Nondetects And Data Analysis for 
environmental data (NADA; \cite{helsel2005}) package for {\tt R}]. However, because more than 50\% of our minor axis data are censored, 
the K-M estimator could not be used to determine the median value of 
the minor axis length, and hence its value was derived using the Maximum Likelihood Estimator (MLE) of the survivor function. 
The mean, median, standard deviation, and 95\% confidence interval of the deconvolved FWHM sizes are given in Table~\ref{table:stat}. 
For example, the median value of the deconvolved $\theta_{\rm maj}$ among the 18 SMGs detected at 3~GHz is $0\farcs54\pm 0\farcs11$ (FWHM), 
and the median major axis FWHM in linear units is $4.2\pm0.9$~kpc as estimated for our SMGs with known redshift (15 sources with either 
a $z_{\rm spec}$ or $z_{\rm phot}$ value available). In the subsequent size analysis we will employ the deconvolved FWHM of the major axis, 
because the value of $\theta_{\rm maj}$ would set the physical extent of a disk-like galaxy, while the minor axis, assuming this simplified disk-like geometry, 
would be given by $\theta_{\rm min}=\theta_{\rm maj}\times \cos(i)$ [defined so that for a disk viewed face-on ($i=0\degr$), $\theta_{\rm min}=\theta_{\rm maj}$].

\begin{figure}[!htb]
\centering
\resizebox{0.9\hsize}{!}{\includegraphics{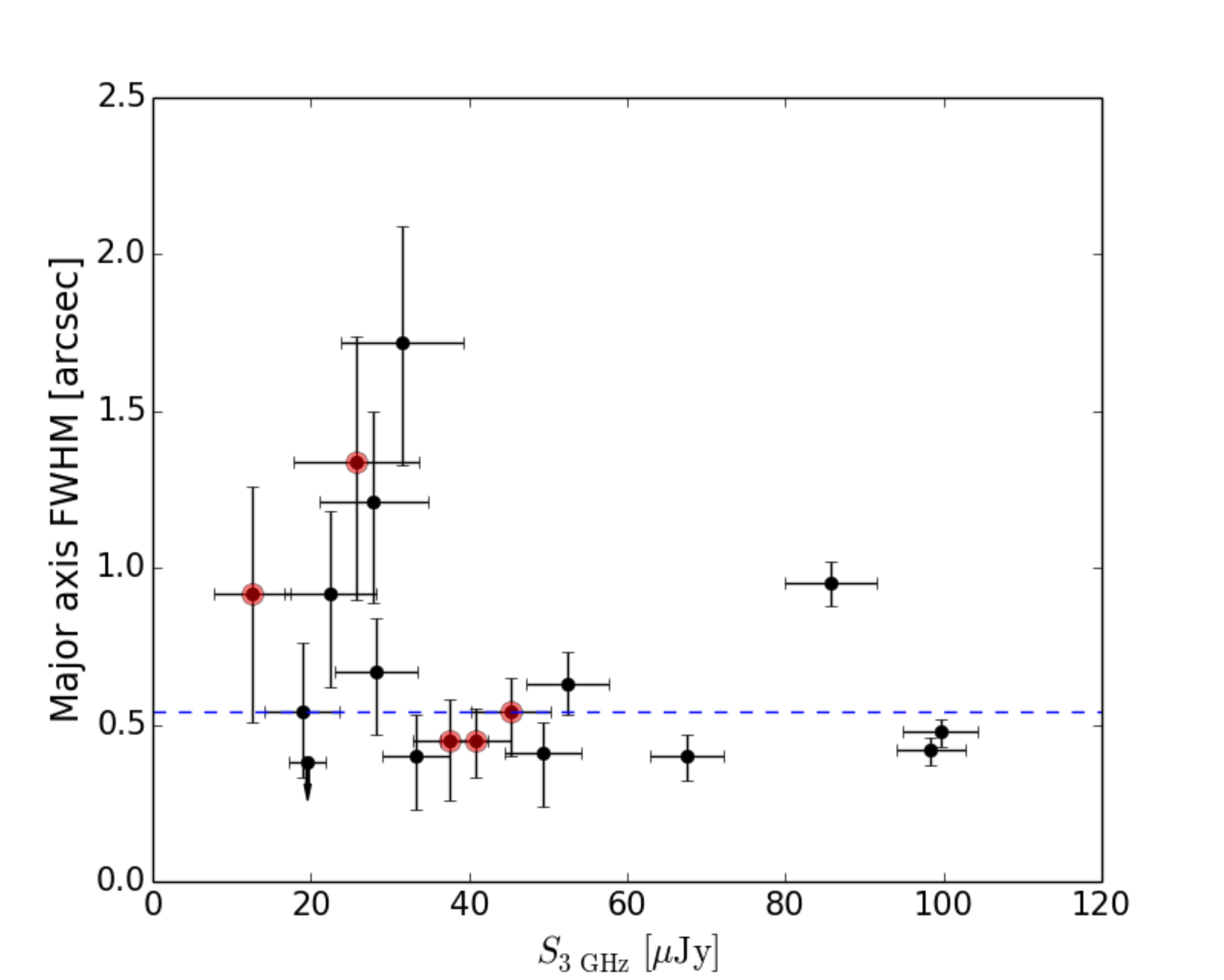}}
\resizebox{0.9\hsize}{!}{\includegraphics{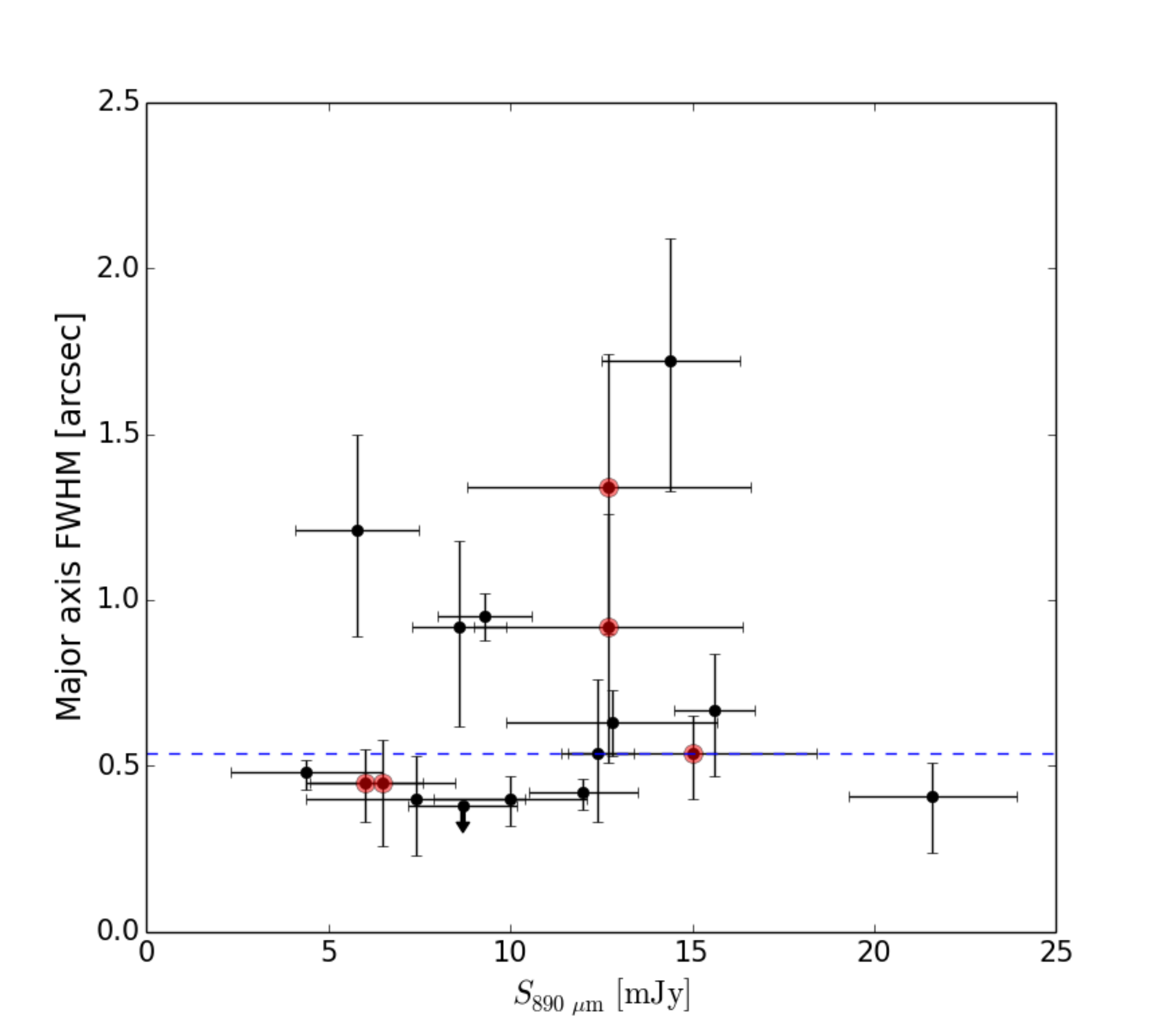}}
\caption{\textbf{Top:} The angular FWHM of the major axis at 3~GHz as a function of the 3~GHz flux density. \textbf{Bottom:} 
Same as above but as a function of the 890~$\mu$m flux density. For AzTEC17a, 19a, 21a, 24b, and 27 the value of $S_{\rm 890\, \mu m}$ was calculated from $S_{\rm 1.3\, mm}$ by 
assuming that the dust emissivity index is $\beta=1.5$; these data points are highlighted by red filled circles in both panels. 
The horizontal dashed line marks the median major axis FWHM of $0\farcs54$. The upper size limit for AzTEC3 is indicated by a downward pointing arrow.}
\label{figure:fluxsize}
\end{figure}

In Fig.~\ref{figure:fluxsize}, we plot the deconvolved major axis FWHM sizes as a function of the 3~GHz flux density (upper panel) and 890~$\mu$m flux density 
(lower panel), where $S_{\rm 890\, \mu m}$ for AzTEC1--15 was taken from Younger et al. (2007, 2009), and for AzTEC17a, 19a, 21a, 24b, and 27 the value of $S_{\rm 890\, \mu m}$ was 
calculated from the 1.3~mm flux density by assuming that $\beta=1.5$ (\cite{miettinen2015}). No statistically significant correlation can be seen 
between the 3~GHz size and the radio or submm flux density, but we note that the largest angular major axis FWHM sizes are preferentially found among the sources with 
the lowest 3~GHz flux densities, although the size uncertainties for those sources are the highest.



\begin{table*}
\caption{Results of Gaussian fits to the 3~GHz sources.}
{\scriptsize
\begin{minipage}{2\columnwidth}
\centering
\renewcommand{\footnoterule}{}
\label{table:results}
\begin{tabular}{c c c c c c c c c c}
\hline\hline 
Source ID & $\alpha_{2000.0}$\tablefootmark{a} & $\delta_{2000.0}$\tablefootmark{a} & $I_{\rm 3\, GHz}$\tablefootmark{b} & $S_{\rm 3\, GHz}$\tablefootmark{b} & S/N & \multicolumn{2}{c}{\underline{FWHM size}\tablefootmark{c}} & P.A.\tablefootmark{c} & Offset \\
       & [h:m:s] & [$\degr$:$\arcmin$:$\arcsec$] & [$\mu$Jy~beam$^{-1}$] & [$\mu$Jy] & & [$\arcsec$] & [kpc] & [$\degr$] & [$\arcsec$]\\ 
\hline
AzTEC1 & 09 59 42.86($\pm0.003$) & +02 29 38.20($\pm0.05$) & $18.3\pm2.4$ & $28.3\pm5.2$ & 8.0 & $0.67^{+0.17}_{-0.20} \times 0.43^{+0.19}_{-0.30}$ & $4.6^{+1.2}_{-1.4} \times 3.0^{+1.3}_{-2.1}$ & $118.1^{+31.8}_{-31.8}$ & 0 \\ [1ex]
AzTEC2\tablefootmark{d} & 10 00 08.04($\pm0.003$) & +02 26 12.26($\pm0.06$) & $15.0\pm2.4$ & $18.9\pm4.7$ & 6.0 & $0.54^{+0.22}_{-0.21} \times < 0.38$ & $4.6^{+1.8}_{-1.8} \times < 3.2$ & $170.9^{\bf +32.6}_{-32.7}$ & 0.16 \\ [1ex]
       & 10 00 08.01($\pm0.007$) & +02 26 10.81($\pm0.08$) & $10.0\pm2.4$ & $14.0\pm5.0$ & 4.3 & $0.72^{+0.31}_{-0.44} \times <0.38$ & \ldots & $72.6^{+28.8}_{-28.8}$ & 1.51 \\ [1ex]
AzTEC3 & 10 00 20.69($\pm0.003$) & +02 35 20.37($\pm0.04$) & $19.6\pm2.3$ & $19.6\pm2.3$ & 8.5 & $<0.38$ & $<2.4$ & \ldots & 0.20  \\ [1ex]
AzTEC4 & 09 59 31.70($\pm0.06$) & +02 30 43.96($\pm0.14$) & $11.4\pm2.3$ & $31.5\pm7.7$ & 5.3 & $1.72^{+0.37}_{-0.39} \times <0.38$ & $11.1^{+2.4}_{-2.5} \times <2.5$ & $150.6^{+8.3}_{-8.3}$ & 0.31  \\ [1ex]
AzTEC5\tablefootmark{d} & 10 00 19.75($\pm0.001$) & +02 32 04.29($\pm0.02$) & $49.2\pm2.4$ & $85.8\pm5.8$ & 21.4 & $0.95^{+0.07}_{-0.07} \times <0.38$ & $7.5^{+0.5}_{-0.6} \times <3.0$ & $41.3^{+4.5}_{-4.5}$ & 0.11  \\ [1ex]
      & 10 00 19.98($\pm0.003$) & +02 32 09.98($\pm0.03$) & $24.0\pm2.4$ & $42.8\pm5.9$ & 10.7 & $1.00^{+0.14}_{-0.16} \times <0.38$ & $8.7^{+1.2}_{-1.4} \times <3.3$ & $81.3^{+8.4}_{-8.4}$ & 6.56 \\ [1ex]
AzTEC6\tablefootmark{e} & 10 00 06.49($\pm0.005$) & +02 38 37.40($\pm0.09$) & $12.3\pm2.3$ & $22.4\pm5.8$ & 5.4 & $0.92^{+0.26}_{-0.30} \times 0.43^{+0.26}_{-0.43}$ & $(6.9^{+1.9}_{-2.3} \times 3.2^{+2.0}_{-3.2})$ & $154.3^{+22.5}_{-22.6}$ & 0.34 \\ [1ex]
AzTEC7 & 10 00 18.06($\pm0.001$) & +02 48 30.43($\pm0.01$) & $89.5\pm2.4$ & $98.4\pm4.4$ & 37.4 & $0.42^{+0.04}_{-0.05} \times < 0.38$ & $3.5^{+0.4}_{-0.4} \times <3.2$ & $29.8^{+6.1}_{-6.2}$ & 0.07  \\ [1ex]
AzTEC8\tablefootmark{d} & 09 59 59.33($\pm0.001$) & +02 34 41.05($\pm0.02$) & $38.8\pm2.4$ & $49.4\pm4.8$ & 16.3 & $0.41^{+0.10}_{-0.17} \times <0.38$ & $3.2^{+0.8}_{-1.3} \times <2.9$ & $46.4^{+0}_{-0}$ & 0.16 \\ [1ex] 
       & 09 59 59.51($\pm0.001$) & +02 34 41.60($\pm0.01$) & $73.5\pm2.4$ & $73.5\pm2.4$ & 29.4 & $<0.38$ & $<3.2$ & \ldots & 2.62 \\[1ex] 
AzTEC9\tablefootmark{f} & 09 59 57.29($\pm0.002$) & +02 27 30.54($\pm0.03$) & $29.4\pm2.2$ & $33.3\pm4.3$ & 13.0 & $\theta_{\rm maj}=0.40^{+0.13}_{-0.17}$ & $\theta_{\rm maj}=3.3^{+1.1}_{-1.4}$ & $33.2^{+24.1}_{-24.1}$ & 0.60 \\ [1ex] 
AzTEC11-N & 10 00 08.90($\pm0.001$) & +02 40 09.52($\pm0.01$) & $57.0\pm2.3$ & $67.5\pm4.6$ & 24.4 & $0.40^{+0.07}_{-0.08} \times <0.38$ & $3.5^{+0.6}_{-0.7} \times <3.3$ & $31.3^{+20.5}_{-20.5}$ & 0.17 \\ [1ex]
AzTEC11-S & 10 00 08.94($\pm0.001$) & +02 40 10.90($\pm0.01$) & $77.5\pm2.3$ & $99.6\pm4.8$ & 33.3 & $0.48^{+0.04}_{-0.05} \times <0.38$ & $4.2^{+0.3}_{-0.5} \times <3.3$ & $163.2^{+14.1}_{-14.1}$ & 1.40 \\ [1ex]
AzTEC12 & 10 00 35.30($\pm0.002$) & +02 43 53.27($\pm0.02$) & $36.5\pm2.5$ & $52.5\pm5.2$ & 14.4 & $0.63^{+0.10}_{-0.10} \times <0.38$ & $5.2^{+0.8}_{-0.8} \times <3.1$ & $78.1^{+14.8}_{-14.8}$ & 0.20 \\ [1ex] 
AzTEC15 & 10 00 12.95($\pm0.006$) & +02 34 34.92($\pm0.10$) & $12.2\pm2.4$ & $27.9\pm6.9$ & 5.4 & $1.21^{+0.29}_{-0.32} \times 0.50^{+0.26}_{-0.50}$ & $9.4^{+2.2}_{-2.5} \times 3.9^{+2.0}_{-3.9}$ & $146.3^{+31.0}_{-15.5}$ & 1.19 \\ [1ex] 
AzTEC17a\tablefootmark{f} & 09 59 39.19($\pm0.001$) & +02 34 03.58($\pm0.02$) & $34.9\pm2.3$ & $40.8\pm4.4$ & 15.1 & $\theta_{\rm maj}=0.45^{+0.10}_{-0.12}$ & $\theta_{\rm maj}=3.5^{+0.8}_{-0.9}$ & $34.4^{+17.1}_{-17.1}$ & 0.26 \\ [1ex] 
AzTEC19a & 10 00 28.72($\pm0.002$) & +02 32 03.68($\pm0.03$) & $31.6\pm2.2$ & $45.3\pm5.0$ & 14.7 & $0.54^{+0.11}_{-0.14} \times 0.44^{+0.21}_{-0.14}$ & $4.2^{+0.8}_{-1.1} \times 3.4^{+1.6}_{-1.1}$ & $174.3^{\bf +44.2}_{-44.3}$ & 0.27  \\ [1ex] 
AzTEC21a & 10 00 02.63($\pm0.01$) & +02 46 42.14($\pm0.10$) & $9.4\pm2.3$ & $25.7\pm7.9$ & 4.2 &  $1.34^{+0.40}_{-0.44} \times 0.67^{+0.32}_{-0.50}$ & $11.0^{+3.3}_{-3.6} \times 5.5^{+2.6}_{-4.1}$ & $95.2^{+21.8}_{-21.8}$ & 1.15  \\ [1ex] 
AzTEC24b\tablefootmark{e} & 10 00 39.28($\pm0.002$) & +02 38 45.14($\pm0.03$) & $28.0\pm2.2$ & $37.6\pm4.7$ & 12.6 & $0.45^{+0.13}_{-0.19} \times 0.43^{+0.15}_{-0.17}$ & $(3.8^{+1.1}_{-1.6} \times 3.6^{+1.3}_{-1.4})$ & $61.3^{+0}_{-0}$ & 2.67  \\ [1ex] 
AzTEC27\tablefootmark{e,f} & 10 00 39.21($\pm0.004$) & +02 40 52.65($\pm0.12$) & $9.9\pm2.3$ & $12.6\pm4.8$ & 4.3 & $\theta_{\rm maj}=0.92^{+0.34}_{-0.41}$ & $(\theta_{\rm maj}=6.4^{+2.4}_{-2.8})$ & $10.1^{+13.0}_{\bf -13.1}$ &
0.47  \\ [1ex] 
\hline 
\end{tabular} 
\tablefoot{The meaning of columns is as follows: (1): SMG name; (2) and (3): peak position of the fitted Gaussian; (4): peak surface brightness; (5): 
total flux density provided by the Gaussian fit; (6): S/N ratio as determined from the maximum pixel value with respect to the rms map noise; 
(7) and (8): deconvolved FWHM size ($\theta_{\rm maj}\times \theta_{\rm min}$) in arcsec and physical kpc; (9) position angle of the fitted Gaussian measured from north through east; 
(10): projected angular offset from the (sub)mm position.\tablefoottext{a}{Formal $1\sigma$ uncertainties in seconds for $\alpha_{2000.0}$ and arcseconds for $\delta_{2000.0}$ 
returned by {\tt JMFIT} are given in parentheses.}\tablefoottext{b}{The quoted error in $I_{\rm 3\, GHz}$ is the $1\sigma$ rms noise in the map determined inside a $\sim300~\sq\arcsec$ box placed near the SMG, and which did not include any 3~GHz sources. The uncertainty in $S_{\rm 3\, GHz}$ represents the formal error determined with {\tt JMFIT}. The uncertainties do not include the absolute calibration uncertainty.}\tablefoottext{c}{The size and P.A. uncertainties represent the minimum/maximum values as returned by {\tt JMFIT}. Note that the P.A. is formally defined to range from $0\degr$ to $180\degr$, but for example for AzTEC2 the maximum P.A. value is $203\fdg5$, which is equivalent to an angle of $203\fdg5-180\degr=23\fdg5$. The minimum and maximum P.A. values for AzTEC8 and 24b are equal to the nominal value, and hence the quoted uncertainties are equal to zero.}\tablefoottext{d}{Two 3 GHz sources were detected. No linear size is reported for the secondary component towards AzTEC2 because of its unknown redshift.}\tablefoottext{e}{For AzTEC6, 24b, and 27 only a lower redshift limit is available (see Table~\ref{table:sample}), and the linear FWHM size quoted in parentheses was calculated at that lower $z$ limit; these linear sizes were not included in the statistical size analysis.}\tablefoottext{f}{For AzTEC9, 17a, and 27 only the major axis could be determined by {\tt JMFIT} (minor axis$=0$).}} 
\end{minipage} }
\end{table*}

\subsection{Spectral index between 1.4 and 3~GHz, and 3~GHz brightness temperature} 

To further characterise the radio continuum properties of our SMGs, we derived their radio spectral index between 
1.4 and 3~GHz ($\alpha_{\rm 1.4\, GHz}^{\rm 3\, GHz}$), and the observed-frame 3~GHz brightness temperature ($T_{\rm B}$). The 1.4~GHz 
flux densities were taken from the COSMOS VLA Deep Catalogue (\cite{schinnerer2010}) for all the sources except AzTEC1, 8, and 11 for 
which $S_{\rm 1.4\, GHz}$ was taken/revised from Younger et al. (2007, 2009); see Col.~(2) in Table~\ref{table:radio}. The angular resolution 
in the 1.4~GHz VLA Deep mosaic is $2\farcs5$ (\cite{schinnerer2010}), while that of the 1.4~GHz VLA-COSMOS Large Project data, used by Younger et al. (2007, 2009), is 
$1\farcs5\times1\farcs4$ (\cite{schinnerer2007}). These are about 3.3 and 1.9 times poorer than in our 3~GHz mosaic, respectively. 
This difference was not taken into account, but we used the 1.4~GHz peak surface brightness as the corresponding source flux density, except for AzTEC8 and 11, 
for which Gaussian-fit based flux densities from Younger et al. (2009) were used (see Table~\ref{table:radio}).
The 1.4 and 3~GHz flux densities were then used to derive $\alpha_{\rm 1.4\, GHz}^{\rm 3\, GHz}$, where we define the spectral 
index as $S_{\nu} \propto \nu^{\alpha}$. The derived spectral indices are listed in Col.~(4) in Table~\ref{table:radio}; 
the quoted errors were propagated from those of the flux densities. The 3~GHz Rayleigh-Jeans brightness temperature was calculated as
$T_{\rm B}=c^2S_{\nu}/(2k_{\rm B}\nu^2 \Omega)$, where $c$ is the speed of light, $k_{\rm B}$ is the Boltzmann constant, and 
the solid angle subtended by the Gaussian source was derived from $\Omega=\pi \theta_{\rm maj}^2/(4 \ln 2)$. The uncertainties in $T_{\rm B}$ were derived 
from those associated with $S_{\rm 3\, GHz}$ and the 3~GHz major axis FWHM size [see Col.~(3) in Table~\ref{table:radio}]. 
We note that Smol{\v c}i{\'c} et al. (2015a) already derived the values of $\alpha_{\rm 1.4\, GHz}^{\rm 3\, GHz}$ for AzTEC1 [$-(0.90\pm0.46)$] 
and AzTEC3 ($>-0.09$). Given the large associated uncertainties, the present spectral index for AzTEC1 [$-(0.69\pm0.61)$] is consistent with the previous value, 
while the lower limit of $\alpha_{\rm 1.4\, GHz}^{\rm 3\, GHz}>-0.91$ we have derived for AzTEC3 is different because of the lower 3~GHz flux density determined here.

\begin{figure}[!h]
\centering
\resizebox{0.9\hsize}{!}{\includegraphics{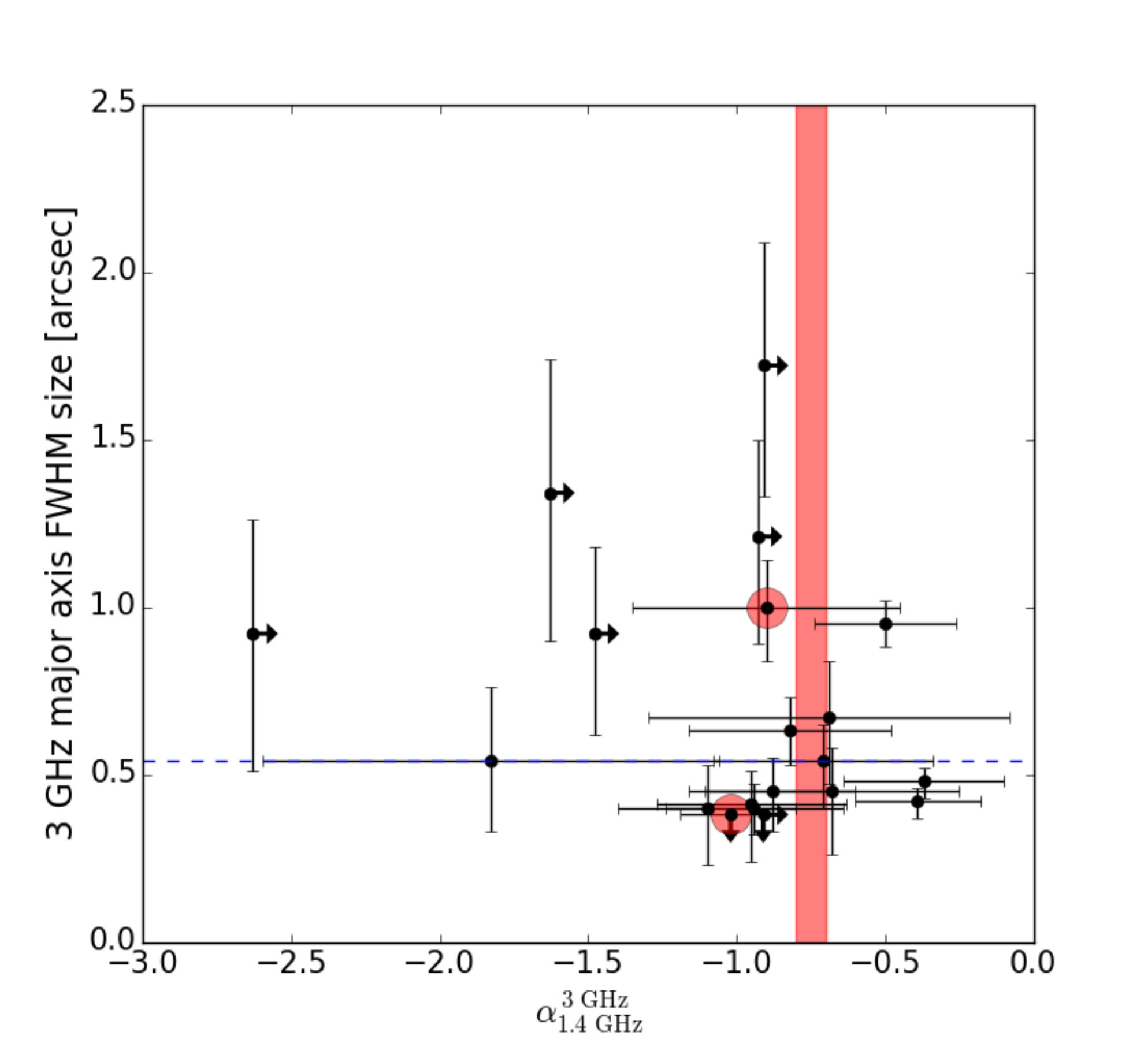}}
\resizebox{0.9\hsize}{!}{\includegraphics{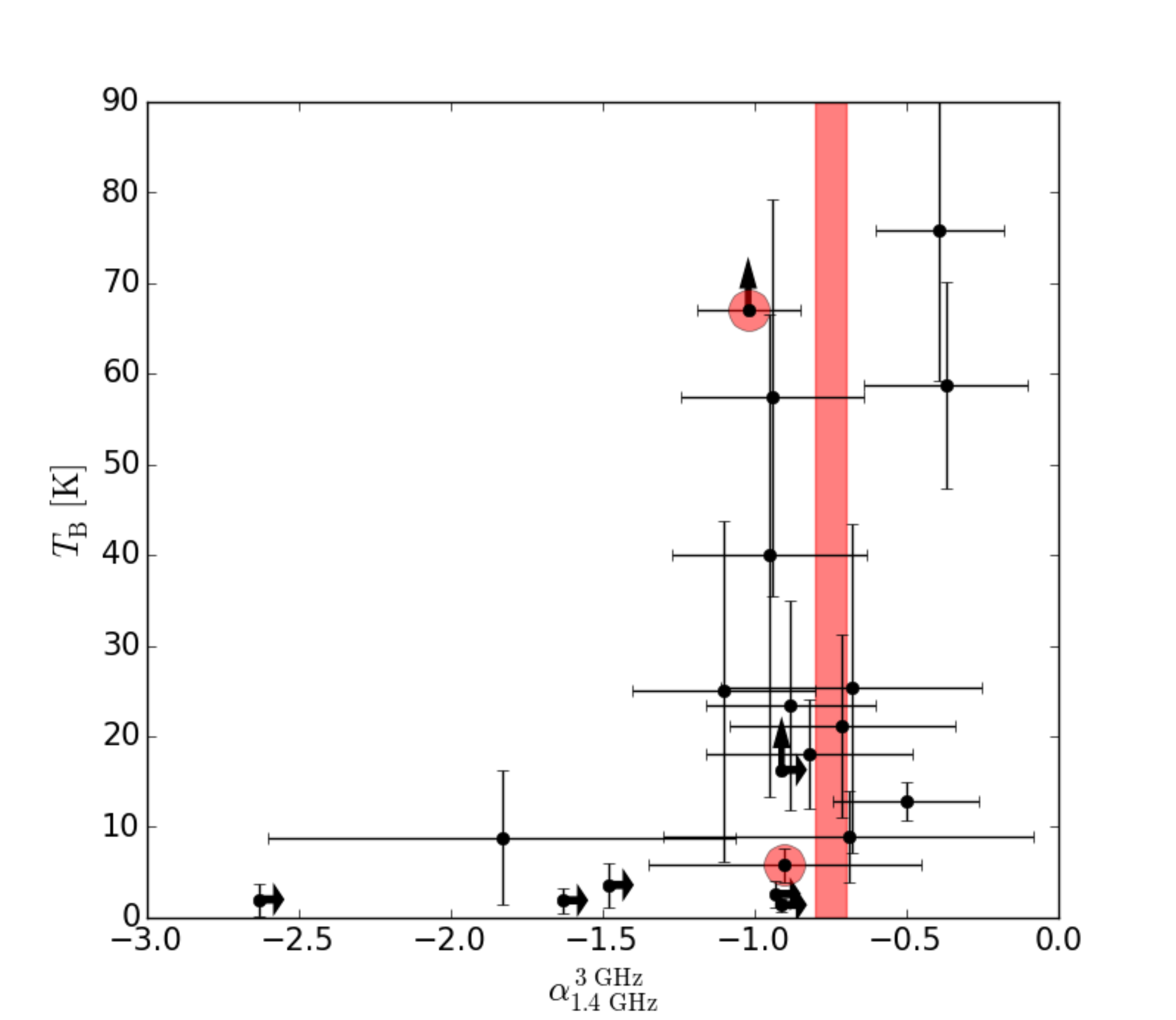}}
\caption{\textbf{Top:} The 3~GHz angular major axis FWHM size plotted against the radio spectral index between the observed frequencies of 1.4 and 3~GHz. 
The arrows pointing right indicate lower limits to $\alpha_{\rm 1.4\, GHz}^{\rm 3\, GHz}$, while the down-pointing arrows show the upper size limits. 
The data points highlighted with red filled circles are for the additional 3~GHz components seen towards AzTEC5 (upper data point) and AzTEC8 (lower data point);  
as discussed in Appendix~A, those are probably not physically related to the SMGs. The horizontal dashed line marks the median major axis FWHM size of our SMGs ($0\farcs54$). 
For reference, the red shaded region shows the radio spectral index range of $-0.8 \ldots -0.7$, 
which is typical of the non-thermal synchrotron radio emission from star-forming galaxies. Within the errors, 11 out of 20 sources 
($55\% \pm 17\% $) shown here have a $\alpha_{\rm 1.4\, GHz}^{\rm 3\, GHz}$ value consistent with this range, and 6 additional sources have a lower limit to 
$\alpha_{\rm 1.4\, GHz}^{\rm 3\, GHz}$ less than $-0.8$. \textbf{Bottom:} The 3~GHz brightness temperature as a function of 
$\alpha_{\rm 1.4\, GHz}^{\rm 3\, GHz}$. The symbols are as in the top panel, except that the arrows pointing up indicate the lower $T_{\rm B}$ 
limits, and the lower and upper red filled circles are for AzTEC5-N and AzTEC8-E, respectively.}
\label{figure:alpha}
\end{figure}

In the top panel of Fig.~\ref{figure:alpha}, we plot the 3~GHz angular major axis FWHM sizes as a function of $\alpha_{\rm 1.4\, GHz}^{\rm 3\, GHz}$, while 
the bottom panel shows the 3~GHz $T_{\rm B}$ values as a function of $\alpha_{\rm 1.4\, GHz}^{\rm 3\, GHz}$. Among local luminous and ultraluminous infrared 
galaxies or (U)LIRGs, it has been found that smaller sources exhibit a flatter radio spectral index (\cite{condon1991}; \cite{murphyetal2013}). 
The observed trend of more compact sources exhibiting flatter radio spectral indices is an indication of increased free-free absorption by ionised gas 
(i.e. free electrons gain energy by absorbing radio photons during their collisions with ions). No such correlation 
is obvious in our data, and the lower $\alpha_{\rm 1.4\, GHz}^{\rm 3\, GHz}$ limits muddy the interpretation. Also, no obvious trend is found
between $T_{\rm B}$ and $\alpha_{\rm 1.4\, GHz}^{\rm 3\, GHz}$ as shown in the bottom panel of Fig.~\ref{figure:alpha}, i.e. sources with a higher $T_{\rm B}$ do not appear to show
spectral flattening (cf.~Fig.~1 in \cite{murphyetal2013}; note their different definition of $S_{\nu} \propto \nu^{-\alpha}$).
The low values of $T_{\rm B}$, ranging from $1.4\pm0.7$~K to $75.9\pm16.6$~K, show that the observed 3~GHz radio emission from our SMGs is powered by star formation activity 
and no evidence of buried AGN activity is visible in our data [AGN have $T_{\rm B}\gtrsim 3.2 \times 10^4$~K at 8.44~GHz ($=\nu_{\rm rest}$ at 
$z\simeq1.8$ for $\nu_{\rm obs}=3$~GHz); \cite{condon1991}; \cite{murphyetal2013}].

\section{Comparison with literature}

\subsection{Previous size measurements of the COSMOS/AzTEC SMGs} 

Besides the present work, the sizes of AzTEC1, 3, 4, 5, and 8 have been previously determined at 3~GHz and/or other 
observed frequencies (see Table~\ref{table:othersizes}). Below, we discuss the size measurements of these five high-redshift SMGs in more detail. 

\begin{figure}[!h]
\centering
\resizebox{0.9\hsize}{!}{\includegraphics{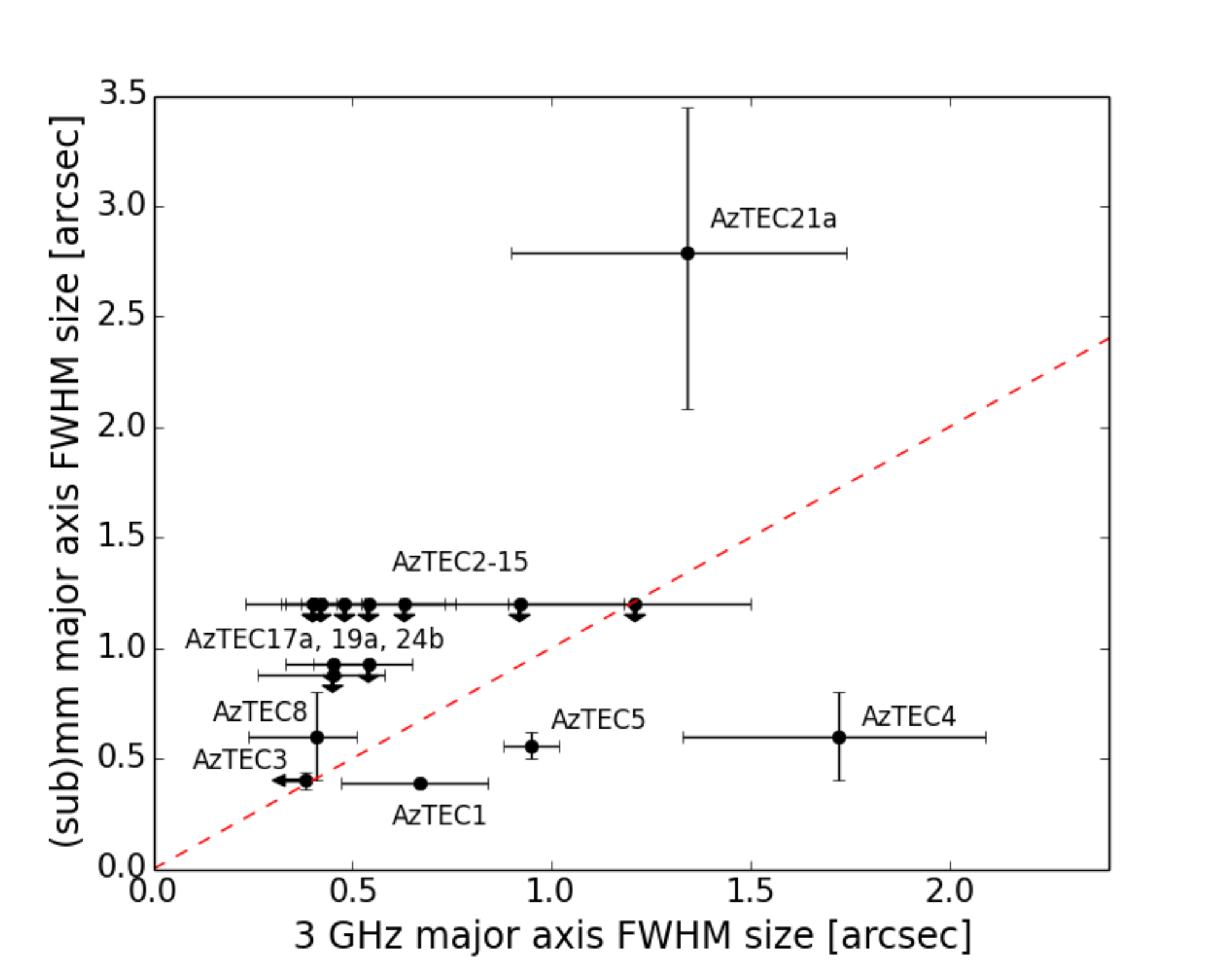}}
\caption{The (sub)mm angular major axis FWHM sizes of our SMGs [at 890~$\mu$m from Younger et al. (2007, 2008, 2009), at 870~$\mu$m from Younger et al. (2010), 
or at 1.3~mm from Miettinen et al. (2015)] plotted against their 3~GHz major axis FWHM sizes. For AzTEC1, 3, 4, 5, 8, and 21a the (sub)mm emission has been (marginally) resolved. 
For AzTEC1, 4, and 8 we plot the Gaussian-fit size from Younger et al. (2008, 2010), while the size of AzTEC3 is that determined from ALMA observations at 1~mm by Riechers et al. (2014). For AzTEC5, the submm size was derived from the 994~$\mu$m ALMA image (O.~Miettinen et al., in prep.) using {\tt JMFIT} (see text for details). 
Similarly, the plotted 1.3~mm size of AzTEC21a was derived by fitting the source using {\tt JMFIT} for a better comparison with the present 3~GHz size. The AzTEC27 data point 
is not shown due to its non-Gaussian shape at 1.3~mm (see text for details). Among AzTEC2--15, the upper 890~$\mu$m size limits -- marked with arrows pointing down -- are $\lesssim1\farcs2$ (\cite{younger2007}, 2009), while the upper limits to the 1.3~mm sizes of AzTEC17a, 19a, and 24b are set to one-half the synthesised beam major axis size at FWHM (\cite{miettinen2015}; Table~2 therein). The red dashed line indicates where the radio and (sub)mm sizes are equal.}
\label{figure:sizecomparison}
\end{figure}

\underline{\textit{AzTEC1}}. We have found that AzTEC1 is resolved ($0\farcs67^{+0.17}_{-0.20} \times 0\farcs43^{+0.19}_{-0.30}$) in our 3~GHz image. 
Smol{\v c}i{\'c} et al. (2015a) employed a 3~GHz submosaic of the COSMOS field, which was based on 130~hr of observations, and had a $1\sigma$ rms noise level of 
4.5~$\mu$Jy~beam$^{-1}$, i.e. about two times higher than in the data used here. AzTEC1 remained unresolved (upper size limit was set to $<0\farcs7$) in 
the previous map although the angular resolution was slightly higher, i.e. $0\farcs7 \times 0\farcs6$. The $0\farcs35\times 0\farcs25$ resolution SMA 890~$\mu$m observations of 
AzTEC1 by Younger et al. (2008) showed the source size to be $\theta_{\rm maj}\times \theta_{\rm min}\sim0\farcs3\times 0\farcs2$ ($\sim0\farcs4\times 0\farcs3$) 
when modelled as a Gaussian (elliptical disk). This Gaussian major axis FWHM at $\lambda_{\rm rest}=167$~$\mu$m is $2.2\pm0.6$ times smaller than 
the size at $\lambda_{\rm rest}=1.9$~cm we have derived (see Fig.~\ref{figure:sizecomparison}). Miettinen et al. (in prep.) used ALMA (PI: A.~Karim) to observe AzTEC1 
at $\lambda_{\rm obs}=870$~$\mu$m ($\lambda_{\rm rest}=163$~$\mu$m) continuum and an angular resolution of $0\farcs30\times 0\farcs29$. 
Fitting the source in the 870~$\mu$m image plane using {\tt JMFIT} results in the deconvolved FWHM size of 
$0\farcs39^{+0.01}_{-0.01} \times 0\farcs31^{+0.01}_{-0.01}$, which is fairly similar to the size derived by Younger et al. (2008) at a comparable wavelength. 
Based on deep UltraVISTA observations ($\sim0\farcs8$ resolution at FWHM), Toft et al. (2014) derived an upper limit of $<2.6$~kpc to the observed-frame NIR size of AzTEC1. 
The authors fit two-dimensional S\'ersic models to the surface brightness distributions, and calculated the effective radius encompassing half the light of the model. 
This size scale corresponds to a Gaussian half width at half maximum (HWHM) size (see Table~1 in \cite{toft2014}), and to be compared with our FWHM diameters 
we multiplied the sizes from Toft et al. (2014) by 2. The physical radius reported by Toft et al. (2014) corresponds to a diameter of $<0\farcs76$ in angular units, 
which suggests that the rest-frame UV-optical size of AzTEC1 could be comparable to its FIR size ($\theta_{\rm maj}\sim0\farcs3$) and/or 1.9~cm radio continuum size 
($ \theta_{\rm maj}=0\farcs67^{+0.17}_{-0.20}$). 

\underline{\textit{AzTEC3}}. This source is unresolved in our 3~GHz image, similarly to that found by Smol{\v c}i{\'c} et al. (2015a) in 
their 3~GHz image (upper size limit was set to $<0\farcs7$). Riechers et al. (2014) used ALMA to observe AzTEC3 
at an angular resolution of $0\farcs63\times 0\farcs56$. In the $\lambda_{\rm obs}=1$~mm ($\lambda_{\rm rest}=159$~$\mu$m) continuum, 
the deconvolved FWHM size of AzTEC3 was derived to be $0\farcs40^{+0.04}_{-0.04} \times 0\farcs17^{+0.08}_{-0.17}$, 
while in the $\lambda_{\rm rest}=158$~$\mu$m $[\ion{C}{II}]$ line emission 
the size was found to be larger, $0\farcs63^{+0.09}_{-0.09} \times 0\farcs34^{+0.10}_{-0.15}$. We have derived the $\lambda_{\rm rest}=1.6$~cm 
upper FWHM size limit of AzTEC3 to be $<0\farcs38$, which is smaller than the aforementioned rest-frame FIR continuum and $[\ion{C}{II}]$ 
sizes (although the major axis FWHM at $\lambda_{\rm rest}=159$~$\mu$m is marginally consistent with our upper size limit; 
see Fig.~\ref{figure:sizecomparison}). The observed-frame NIR diameter of AzTEC3 derived by Toft et al. (2014) is $<4.8$~kpc, 
i.e. $<0\farcs76$, which is also consistent with our radio emission FWHM size, and with the FIR FWHM size from Riechers et al. (2014).  

\underline{\textit{AzTEC4}}. For this source, the FWHM size at 3~GHz is determined to be $1\farcs72^{+0.37}_{-0.39} \times < 0\farcs38$. The source appears elongated with a major-to-minor axis ratio of 
$>3.5$, but as shown in Fig.~\ref{figure:maps}, the 3~GHz peak position is not well determined by {\tt JMFIT}. The $0\farcs86\times 0\farcs77$ resolution SMA 870~$\mu$m observations of 
AzTEC4 by Younger et al. (2010) showed the source size to be $\theta_{\rm maj}\times \theta_{\rm min}=(0\farcs6 \pm 0\farcs2)\times (0\farcs4 \pm 0\farcs2)$ [$(1\farcs0 \pm 0\farcs4)\times (0\farcs7 \pm 0\farcs6)$] 
when modelled as a Gaussian (elliptical disk). This Gaussian major axis FWHM at $\lambda_{\rm rest}=147$~$\mu$m is $2.9^{+2.4}_{-1.2}$ times smaller than the size at $\lambda_{\rm rest}=1.7$~cm we have derived 
(see Fig.~\ref{figure:sizecomparison}). The observed-frame NIR diameter of AzTEC4 derived by Toft et al. (2014) is $<5.0$~kpc ($<0\farcs78$), which is consistent with the Gaussian FWHM at rest-frame FIR derived by 
Younger et al. (2010).

\underline{\textit{AzTEC5}}. The 3~GHz FWHM size we have determined for this source is $0\farcs95^{+0.07}_{-0.07} \times < 0\farcs38$, i.e. the major axis is resolved while the minor axis is unresolved. 
Miettinen et al. (in prep.) used ALMA (PI: A.~Karim) to observe AzTEC5 at $\lambda_{\rm obs}=994$~$\mu$m ($\lambda_{\rm rest}\simeq245$~$\mu$m) 
continuum and an angular resolution of $0\farcs52\times 0\farcs30$. The source was resolved into two components with a projected separation of $0\farcs75$ ($\sim5.86$~kpc at the source redshift). The northern ALMA component is perfectly coincident with our 3~GHz source ($0\farcs03$ offset), but we note that the 3~GHz emission extends towards the southern ALMA FIR component, and hence the detected 3~GHz emission encompasses the two ALMA-detected components. Fitting the ALMA sources in the image plane using {\tt JMFIT} results in the deconvolved FWHM size of $0\farcs45^{+0.04}_{-0.02} \times 0\farcs28^{+0.04}_{-0.08}$ for the northern component, and $0\farcs56^{+0.06}_{-0.06} \times 0\farcs38^{+0.08}_{-0.09}$ for the southern component. The major axis FWHM length of $0\farcs95^{+0.07}_{-0.07}$ at 3~GHz is comparable to the sum of the major axes of the ALMA emission from the two sources ($\simeq1\arcsec$). Toft et al. (2014) used high resolution (FWHM$\sim0\farcs2$) data from the Wide Field Camera 3 (WFC3) aboard the \textit{Hubble Space Telescope} to determine the rest-frame UV/optical size of AzTEC5. The diameter derived from their reported radius is $1.0\pm0.8$~kpc ($0\farcs12\pm0\farcs10$). 
The major axis FWHM of AzTEC5 at 3~GHz is $7.9\pm6.6$ times larger than its UV/optical diameter.

\underline{\textit{AzTEC8}}. For this source, the FWHM size at 3~GHz is determined to be $0\farcs41^{+0.10}_{-0.17} \times < 0\farcs38$. The major axis is only marginally resolved, while the minor axis is unresolved. The $0\farcs86\times 0\farcs55$ resolution SMA 870~$\mu$m observations of AzTEC8 by Younger et al. (2010) showed the source size to be $\theta_{\rm maj}\times \theta_{\rm min}=(0\farcs6 \pm 0\farcs2)\times (0\farcs5 \pm 0\farcs3)$ [$(1\farcs0 \pm 0\farcs5)\times (0\farcs4 \pm 0\farcs8)$] when modelled as a Gaussian (elliptical disk). This Gaussian major axis FWHM at $\lambda_{\rm rest}=208$~$\mu$m is $1.5^{+1.8}_{-0.7}$ times larger than the radio size at $\lambda_{\rm rest}=2.4$~cm we have derived (see Fig.~\ref{figure:sizecomparison}). The observed-frame NIR diameter of AzTEC8 derived by Toft et al. (2014), $<6.0$~kpc ($<0\farcs78$), is consistent with our 3~GHz size and the Gaussian rest-frame FIR size determined by Younger et al. (2010).

For the remaining of our SMGs mostly upper size limits at other wavelengths are available. Younger et al. (2007, 2009) constrained the observed-frame 890~$\mu$m sizes of 
AzTEC1--15 to $\lesssim1\farcs2$ (i.e. the sources were unresolved), with the exception of AzTEC11 that was found to be resolved but best modelled as a double point source.
Of the 3~GHz-detections among AzTEC16--30, all the other SMGs except AzTEC21a were found to be unresolved by Miettinen et al. (2015) in the $\sim1\farcs8$ resolution PdBI 1.3~mm images. A Gaussian fit to AzTEC21a yielded a rather poorly constrained deconvolved FWHM of 
$(2\farcs6\pm 1\farcs2)  \times (0\farcs3\pm 0\farcs5)$. For a fair comparison with the present 3~GHz size, we fitted the source using {\tt JMFIT}, and obtained 
a 1.3~mm FWHM size of $2\farcs79^{+0.66}_{-0.71} \times 0\farcs60^{+0.59}_{-0.60}$ (${\rm P.A.}=47\fdg8^{+11.0}_{-15.3}$), which is comparable to the 
aforementioned value, and the major axis is about $2.1^{+1.7}_{-0.9}$ times larger than that at  3~GHz ($1\farcs34^{+0.40}_{-0.44}$). 
AzTEC21a is potentially a blend of smaller (sub)mm-emitting sources (\cite{miettinen2015}), hence appears more extended at 1.3~mm than its radio size. 
We note that the PdBI 1.3~mm emission of AzTEC27 could not be well modelled by a single Gaussian source model; 
the major axis FWHM was determined to be $\theta_{\rm maj}=3\farcs6$ (\cite{miettinen2015}). Similarly to AzTEC21a, AzTEC27 could be a blend of more compact sources. 
Higher-resolution (sub)mm imaging is required to examine the possible multiplicity of AzTEC21a and 27. 
The constraints on the (sub)mm FWHM sizes of our SMGs are listed in Table~\ref{table:othersizes} and plotted against their 3~GHz major axis FWHM sizes in Fig.~\ref{figure:sizecomparison}. 

Besides for AzTEC1, 3, 4, 5, and 8, Toft et al. (2014) also derived the rest-frame UV/optical sizes for AzTEC10 and AzTEC15 (see their Table~1). 
The measurements were based on Ultra\-VISTA observations. The diameters were found to be $1.4\pm0.2$~kpc ($0\farcs18^{+0.02}_{-0.04}$) for AzTEC10, 
and $10.0\pm1.6$~kpc ($1\farcs28^{+0.22}_{-0.20}$) for AzTEC15. The 3~GHz major axis FWHM of AzTEC15 ($\theta_{\rm maj}=1\farcs21^{+0.29}_{-0.32}$) is 
in good agreement with its UV/optical extent, while AzTEC10 was not detected at 3~GHz. 
We note that heavy obscuration by dust can lead to an apparent compact size at the rest-frame UV/optical wavelengths. However, one would expect the central galactic regions to be more extincted compared to the outer portions, which could affect the surface brightness profile in such a way that the measured size (e.g. the half-light radius) is larger than in the case of no differential dust extinction. However, if the extincted outer parts of a galaxy fall below the detection limit, the effect might go in the opposite direction.

\begin{table}
\renewcommand{\footnoterule}{}
\caption{The 3~GHz size distribution statistics.}
{\small
\begin{minipage}{1\columnwidth}
\centering
\label{table:stat}
\begin{tabular}{c c}
\hline\hline 
Parameter & Value\tablefootmark{a} \\
\hline
Mean $\theta_{\rm maj}$ & $0\farcs71\pm0\farcs09$ ($5.5\pm0.7$~kpc) \\
Median $\theta_{\rm maj}$ & $0\farcs54\pm0\farcs11$ ($4.2\pm0.9$~kpc) \\
Mean $\theta_{\rm min}$ & $0\farcs45\pm0\farcs02$ ($3.3\pm0.2$~kpc) \\
Median $\theta_{\rm min}$\tablefootmark{b} & $0\farcs35\pm0\farcs04$ ($2.3\pm0.4$~kpc) \\
Standard deviation of $\theta_{\rm maj}$ & $0\farcs39$ (2.9~kpc) \\ 
Standard deviation of $\theta_{\rm min}$ & $0\farcs07$ (0.8~ kpc)\\
95\% confidence interval of $\theta_{\rm maj}$\tablefootmark{c} & $0\farcs54-0\farcs89$ (4.0--6.9~kpc)\\ 
95\% confidence interval of $\theta_{\rm min}$\tablefootmark{c} & $0\farcs42-0\farcs49$ (2.9--3.7~kpc)\\ 
\hline 
\end{tabular} 
\tablefoot{\tablefoottext{a}{The sample size of the major (minor) axis angular sizes is 18 (15), while that of the linear sizes is 15 (13), i.e. the number of SMGs with either 
a $z_{\rm spec}$ or $z_{\rm phot}$ value available.}\tablefoottext{b}{The median value of $\theta_{\rm min}$ could not be derived using the K-M estimator. Hence, it was calculated using the MLE (assuming lognormal distribution), which is almost identical to the K-M function.}\tablefoottext{c}{A two-sided 95\% confidence interval for the mean value computed using the K-M method.}}
\end{minipage} 
}
\end{table}

\begin{table}
\renewcommand{\footnoterule}{}
\caption{Radio continuum characteristics of the 3~GHz detections.}
{\small
\begin{minipage}{1\columnwidth}
\centering
\label{table:radio}
\begin{tabular}{c c c c}
\hline\hline 
Source ID & $S_{\rm 1.4\, GHz}$\tablefootmark{a} & $T_{\rm B}$\tablefootmark{b} & $\alpha_{\rm 1.4\, GHz}^{\rm 3\, GHz}$\tablefootmark{c}  \\
          & [$\mu$Jy] & [K] & \\
\hline
AzTEC1 & $48\pm12$ & $8.9\pm5.0$ & $-(0.69\pm0.61)$ \\ 
AzTEC2 & $76\pm14$ & $8.8\pm7.4$ & $-(1.83\pm0.77)$ \\ 
AzTEC3 & $<30$ & $>16.3$ & $>-0.91$\\ 
AzTEC4 & $<36$ & $1.4\pm0.7$ & $>-0.91$ \\ 
AzTEC5 & $126\pm15$ & $12.9\pm2.1$ & $-(0.50\pm0.24)$ \\ 
AzTEC5-N & $85\pm15$ & $5.8\pm1.9$ & $-(0.90\pm0.45)$ \\
AzTEC6 & $<38$ & $3.6\pm2.4$ & $>-1.48$ \\ 
AzTEC7 & $132\pm22$ & $75.9\pm16.6$ & $-(0.39\pm0.21)$ \\ 
AzTEC8-W & $102\pm13$ & $40.0\pm26.6$ & $-(0.95\pm0.32)$ \\ 
AzTEC8-E & $160\pm23$ & $>67$ & $-(1.02\pm0.17)$ \\ 
AzTEC9 & $68\pm13$ & $25.0\pm18.8$ & $-(1.10\pm0.30)$ \\ 
AzTEC11-N & $138\pm26$ & $57.4\pm21.9$ & $-(0.94\pm0.30)$\\ 
AzTEC11-S & $132\pm26$ & $58.8\pm11.4$ & $-(0.37\pm0.27)$\\ 
AzTEC12 & $98\pm16$ & $18.0\pm6.0$ & $-(0.82\pm0.34)$ \\
AzTEC15 & $<32$ & $2.6\pm1.5$ & $>-0.93$ \\ 
AzTEC17a & $68\pm13$ & $23.4\pm11.6$ & $-(0.88\pm0.28)$\\ 
AzTEC19a & $78\pm12$ & $21.1\pm10.1$ & $-(0.71\pm0.37)$\\ 
AzTEC21a & $<44$ & $1.9\pm1.4$ & $>-1.63$\\
AzTEC24b & $63\pm13$ & $25.3\pm18.2$ & $-(0.68\pm0.43)$\\ 
AzTEC27 & $<39$ & $2.0\pm1.8$ & $>-2.63$  \\
\hline 
\end{tabular} 
\tablefoot{\tablefoottext{a}{The values of $S_{\rm 1.4\, GHz}$ were taken from the COSMOS VLA Deep Catalogue May 2010 
(\cite{schinnerer2010}) except for AzTEC1, 8, and 11. AzTEC1 exhibits $4\sigma$ 1.4~GHz emission, hence is not listed in the VLA catalogue, which is comprised of $\geq5\sigma$ sources. The value of $S_{\rm 1.4\, GHz}$ for AzTEC1 was taken as the peak surface brightness multiplied by 1.15 to correct for BWS (\cite{younger2007}). The resulting flux density of $48\pm12$~$\mu$Jy agrees with the value reported by Younger et al. (2007, Table~2 therein). For AzTEC8 and 11 we adopted the 1.4~GHz flux densities from Younger et al. (2009, Table~2 therein), and multiplied them by 1.15 to correct for BWS as noted by the authors. The COSMOS VLA catalogue values of $S_{\rm 1.4\, GHz}$ for AzTEC8-W and AzTEC8-E are $237\pm52$~$\mu$Jy and 186~$\mu$Jy (no error given), respectively. However, AzTEC8-E is a stronger 1.4~GHz source than AzTEC8-W (\cite{younger2009}). For AzTEC11 the COSMOS VLA catalogue gives an integrated 
flux density value of $S_{\rm 1.4\, GHz}=302\pm45$~$\mu$Jy. Hence, we adopted the $S_{\rm 1.4\, GHz}$ values for AzTEC11-N and 11-S from Younger et al. (2009). The $3\sigma$ upper limits are reported for the non-detections, where $1\sigma\simeq10-14.7$~$\mu$Jy~beam$^{-1}$.}\tablefoottext{b}{The value of $T_{\rm B}$ refers to the Rayleigh-Jeans brightness temperature at $\nu_{\rm obs}=3$~GHz.}\tablefoottext{c}{Radio spectral index between the observed-frame frequencies of 1.4~GHz and 3~GHz.}  }
\end{minipage} 
}
\end{table}

\begin{table}
\renewcommand{\footnoterule}{}
\caption{Rest-frame FIR/submm and UV/optical sizes of our 3~GHz detected SMGs. Besides the angular sizes, 
the physical sizes are given in parentheses when a spectroscopic or photometric redshift is available.}
{\scriptsize
\begin{minipage}{1\columnwidth}
\centering
\label{table:othersizes}
\begin{tabular}{c c c}
\hline\hline 
Source ID & FIR/submm size\tablefootmark{a} & UV/opt. size\tablefootmark{b} \\
\hline
AzTEC1 & $0\farcs3\times 0\farcs2$\tablefootmark{c} & $<0\farcs76$ ($<5.2$~kpc) \\
       & ($2.1\times1.4$~kpc$^2$)\tablefootmark{c} & \ldots \\  
       & $0\farcs4\times 0\farcs3$\tablefootmark{c} & \ldots \\
       & ($2.8\times2.1$~kpc$^2$)\tablefootmark{c} & \ldots \\  
       & $0\farcs39^{+0.01}_{-0.01} \times 0\farcs31^{+0.01}_{-0.01}$\tablefootmark{d}  & \ldots \\
       & ($2.7^{+0.1}_{-0.1} \times 2.1^{+0.1}_{-0.1}$~kpc$^2$)\tablefootmark{d} & \ldots \\  
AzTEC2 & $\lesssim1\farcs2$ ($\lesssim10.1$~kpc) & \ldots \\
AzTEC3 & $0\farcs40^{+0.04}_{-0.04} \times 0\farcs17^{+0.08}_{-0.17}$\tablefootmark{e} & $<0\farcs76$ ($<4.8$~kpc) \\
       & ($2.5^{+0.3}_{-0.3} \times 1.1^{+0.5}_{-1.1}$~kpc$^2$)\tablefootmark{e} & \ldots \\  
AzTEC4 & $(0\farcs6 \pm 0\farcs2)\times (0\farcs4 \pm 0\farcs2)$\tablefootmark{f} & $<0\farcs78$ ($<5.0$~kpc) \\
       & [$(3.9\pm1.3) \times (2.6\pm1.3)$~kpc$^2$]\tablefootmark{f} & \ldots \\ 
       & $(1\farcs0 \pm 0\farcs4)\times (0\farcs7 \pm 0\farcs6)$\tablefootmark{f} & \ldots \\ 
       & [$(6.5\pm2.6) \times (4.5\pm3.9)$~kpc$^2$]\tablefootmark{f} & \ldots \\
AzTEC5 & $0\farcs45^{+0.04}_{-0.02} \times 0\farcs28^{+0.04}_{-0.08}$\tablefootmark{g} & $0\farcs12\pm0\farcs10$ ($1.0\pm0.8$~kpc) \\
       & ($3.5^{+0.3}_{-0.1} \times 2.2^{+0.3}_{-0.6}$~kpc$^2$)\tablefootmark{g} & \ldots \\
AzTEC6 & $\lesssim1\farcs2$ & \ldots \\
AzTEC7 & $\lesssim1\farcs2$ ($\lesssim10.1$~kpc) & \ldots \\
AzTEC8 & $(0\farcs6 \pm 0\farcs2)\times (0\farcs5 \pm 0\farcs3)$\tablefootmark{f} & $<0\farcs78$ ($<6.0$~kpc)\\
       & [$(4.6\pm1.5) \times (3.9\pm2.3)$~kpc$^2$]\tablefootmark{f} & \ldots \\ 
       & $(1\farcs0 \pm 0\farcs5)\times (0\farcs4 \pm 0\farcs8)$\tablefootmark{f} & \ldots \\ 
       & [$(7.7\pm3.9) \times (3.1\pm6.2)$~kpc$^2$]\tablefootmark{f} & \ldots \\
AzTEC9 & $\lesssim1\farcs2$ ($\lesssim10.0$~kpc) & \ldots \\
AzTEC11-N & $\lesssim1\farcs2$ ($\lesssim10.4$~kpc) & \ldots \\
AzTEC11-S & $\lesssim1\farcs2$ ($\lesssim10.4$~kpc) & \ldots \\
AzTEC12 & $\lesssim1\farcs2$ ($\lesssim9.9$~kpc) & \ldots \\
AzTEC15 & $\lesssim1\farcs2$ ($\lesssim9.3$~kpc) & $1\farcs28^{+0.22}_{-0.20}$ ($10.0\pm1.6$~kpc)\\
AzTEC17a & $\lesssim0\farcs93$ ($\lesssim7.3$~kpc) & \ldots \\
AzTEC19a & $\lesssim0\farcs93$ ($\lesssim7.2$~kpc) & \ldots \\
AzTEC21a & $2\farcs79^{+0.66}_{-0.71} \times 0\farcs60^{+0.59}_{-0.60}$\tablefootmark{h} & \ldots \\
         & ($22.9^{+5.4}_{-5.8} \times 4.9^{+4.8}_{-4.9}$~kpc$^2$)\tablefootmark{h} & \ldots  \\  
AzTEC24b & $\lesssim0\farcs88$ & \ldots \\
AzTEC27 & $\theta_{\rm maj}=3\farcs6$\tablefootmark{i} & \ldots \\
\hline 
\end{tabular} 
\tablefoot{\tablefoottext{a}{The upper FWHM size limits for AzTEC2--15 at observed-frame 890~$\mu$m are from Younger et al. (2007, 2009), while those for AzTEC17a, 19a, and 24b refer to observed-frame 1.3~mm (\cite{miettinen2015}) and represent half of the beam major axis FWHM.}\tablefoottext{b}{The diameter at rest-frame UV/optical derived from the effective radii from Toft et al. (2014) (see text for details).}\tablefoottext{c}{The FWHM size derived from SMA 890~$\mu$m data by Younger et al. (2008) when modelling the source as a Gaussian (upper value) or elliptical disk (lower value).}\tablefoottext{d}{The FWHM size measured from the ALMA 870~$\mu$m image (O.~Miettinen et al., in prep.) using {\tt JMFIT}.}\tablefoottext{e}{A deconvolved FWHM size derived through ALMA 1~mm observations by Riechers et al. (2014).}\tablefoottext{f}{The FWHM size derived from SMA 870~$\mu$m data by Younger et al. (2010) when modelling the source as a Gaussian (upper value) or elliptical disk (lower value).}\tablefoottext{g}{The FWHM size measured from the ALMA 994~$\mu$m image (O.~Miettinen et al., in prep.) using {\tt JMFIT}.}\tablefoottext{h}{The observed-frame 1.3~mm FWHM size of AzTEC21a derived here using {\tt JMFIT}.}\tablefoottext{i}{The major axis FWHM at 1.3~mm from Miettinen et al. (2015).}            }
\end{minipage} 
}
\end{table}

\subsection{Comparison to SMG sizes from the literature} 

In this subsection, we discuss the SMG sizes derived in previous surveys at different wavelengths. The measured sizes discussed below are derived using the following 
four observational probes: \textit{i)} radio continuum emission at centimetre wavelengths; \textit{ii)} (sub)mm continuum emission (typically corresponding to 
rest-frame FIR); \textit{iii)} molecular spectral line emission arising from rotational transitions of CO; and \textit{iv)} rest-frame optical emission tracing 
the spatial extent of the stellar content. A selection of size distributions derived from the reported data in the studies discussed below is shown in 
Fig.~\ref{figure:sizes} alongside with our 3~GHz size distribution. 

\subsubsection{Radio sizes}

A previous work of immediate interest for comparison with our results is the MERLIN/VLA 1.4~GHz survey ($1\sigma=6$~$\mu$Jy~beam$^{-1}$; 
$\sim0\farcs52 \times 0\farcs48$ resolution) by Biggs \& Ivison (2008) of the Lockman Hole SMGs (spanning a redshift range of $z_{\rm spec}=1.147-2.689$). 
The median deconvolved FWHM size we derived from their data (their Table~3 of AIPS/{\tt JMFIT}-derived sizes) is 
$(0\farcs61\pm0\farcs10)\times(0\farcs31\pm0\farcs08)$.\footnote{The median sizes from other works that we report in this section were derived (when relevant) 
using a survival analysis as described in Sect.~4.1.} This is comparable to our median size of $(0\farcs54\pm 0\farcs11)\times(0\farcs35\pm0\farcs04)$ derived 
from 2.1 times higher frequency observations. The median linear size from Biggs \& Ivison (2008), $(6.3\pm0.9)\times(3.3\pm0.7)$~kpc$^2$ (scaled to the \textit{Planck} 2015 cosmology), is $1.5\pm0.4$ times larger in the major axis than our value of $(4.2\pm0.9)\times(2.3\pm0.4)$~kpc$^2$. If we consider only those SMGs in our sample that lie in the redshift range studied by Biggs \& Ivison (2008), i.e. AzTEC7, 11-N, 11-S, and 12, the discrepancy becomes more significant: the median linear major axis of these four sources is $3.5\pm0.5$~kpc, i.e. $1.8\pm0.4$ times smaller than that from Biggs \& Ivison (2008). However, given the relatively small number of sources in these two (sub)samples [$z_{\rm spec}$ values are available for eight SMGs in the Biggs \& Ivison (2008) sample], the latter comparison might be susceptible to small number statistics.

To see how the millimetre flux densities of the Biggs \& Ivison (2008) SMGs compare to those of our SMGs, we compiled their Bolocam 1.1~mm (\cite{laurant2005}), MAMBO 1.2~mm (\cite{ivison2005}), and SCUBA 850~$\mu$m (\cite{ivison2007}) flux densities and converted them to 1.1~mm flux densities assuming that $\beta=1.5$ when needed. The resulting flux density range is $S_{\rm 1.1\, mm}=1.4_{-0.7}^{+0.6}-6.0_{-1.4}^{+1.4}$~mJy, which includes somewhat fainter sources than ours with the JCMT/AzTEC 1.1~mm flux densities of $S_{\rm 1.1\, mm}=3.3_{-1.6}^{+1.4}-9.3_{-1.3}^{+1.3}$~mJy (\cite{scott2008}). However, the two 1.1~mm flux density ranges are comparable within the uncertainties, and hence the radio size comparison is reasonable. Moreover, we found no correlation between the Biggs \& Ivison (2008) SMGs' radio sizes and their millimetre flux densities (cf.~our Fig.~\ref{figure:fluxsize}, bottom panel). We note that Chapman et al. (2004), who studied 12 Hubble Deep Field SMGs ($z=1.01-2.91$) using the MERLIN/VLA 1.4~GHz observations ($0\farcs2-0\farcs3$ resolution), found that in most cases ($67\pm24\%$) the radio emission is resolved on angular scales of $\sim 1\arcsec$ ($\sim 8.5$~kpc at their median redshift of $z=2.2$). The median diameter (measured above $3\sigma$ emission) was reported to be $0\farcs83\pm0\farcs14$ ($7.0^{+1.2}_{-1.4}$~kpc), which is larger than our median 3~GHz major axis FWHM by a factor of $1.5\pm0.4$, although it was not specified whether the median diameter refers to the deconvolved FWHM as determined in the present paper [we note that according to Biggs \& Ivison (2008), the size reported by Chapman et al. (2004) is the largest extent within the $3\sigma$ contour, hence not directly comparable with our FWHM sizes]. The authors concluded that their SMGs are extended starbursts and therefore different from local ULIRGs with sub-kpc nuclear starburst regions (e.g. \cite{condon1991}). Biggs et al. (2010) used 18~cm Very Long Baseline Interferometry (VLBI) radio observations at a very high angular resolution of about 30~mas to examine the sizes of a sample of six compact SMGs drawn from the Biggs \& Ivison (2008) sample. Only two of these six SMGs ($33\pm6\%$) were found to host an ultra-compact AGN radio core, and the authors concluded that the radio emission from their SMGs is mostly arising from star formation rather than from an AGN activity.

\subsubsection{(Sub)mm sizes}

Ikarashi et al. (2015) recently derived a size distribution for a sample of 13 high-redshift ($z_{\rm phot}\sim3-6$) SMGs through 1.1~mm ALMA observations at $\sim 0\farcs2$ resolution. 
Their SMGs were originally discovered in the ASTE/AzTEC 1.1~mm observations of the Subaru/\textit{XMM-Newton} Deep Field (S.~Ikarashi et al., in prep.), and the reported ALMA 1.1~mm flux 
densities lie in the range of $S_{\rm 1.1\, mm}=(1.23\pm0.07)-(3.45\pm0.10)$~mJy. Compared to the JCMT/AzTEC 1.1~mm flux densities of our SMGs, namely 
$S_{\rm 1.1\, mm}=3.3_{-1.6}^{+1.4}-9.3_{-1.3}^{+1.3}$~mJy (\cite{scott2008}), the Ikarashi et al. (2015) SMGs are fainter, which hampers the direct comparison of these two SMG samples. 
 Ikarashi et al. (2015) measured the sizes using the $uv$ visibility data directly, and assumed symmetric Gaussian profiles. From the values given in their Table~1 we derived a median FWHM 
of $0\farcs22\pm0\farcs04$. The linear radii reported by the authors are very compact, $\sim0.3-1.3$~kpc (median $\sim0.8$~kpc), which translate to diameters of $\sim0.6-2.6$~kpc (median $\sim1.6$~kpc). 
These sizes suggest that the high-redshift ($z\gtrsim3$) SMGs are associated with a compact starburst region (as seen at $\lambda_{\rm rest}<160-289$~$\mu$m), 
and Ikarashi et al. (2015) concluded that the median SFR surface density of their SMGs, $\sim10^2$~M$_{\sun}$~yr$^{-1}$~kpc$^{-2}$, is comparable to that of local merger-driven (U)LIRGs and higher than those of low- and high-$z$ (extended) disk galaxies.

Simpson et al. (2015a) carried out a high-resolution ($0\farcs35\times 0\farcs25$) 870~$\mu$m ALMA survey of a sample of 30 of the brightest 850~$\mu$m-selected SMGs from 
the SCUBA-2 Cosmology Legacy Survey of the UKIDSS Ultra Deep Survey (UDS) field. Their target SMGs have 850~$\mu$m flux densities 
of $S_{\rm 850\, \mu m}=8-16$~mJy ($>4\sigma$; $S_{\rm 1.1\, mm}=3.2-6.5$~mJy if $\beta=1.5$), and hence are mostly comparable to our flux-limited sample (only three of our SMGs lie above this flux density range). For a subsample of 23 SMGs (detected at ${\rm S/N}_{\rm 870\, \mu m}>10$), they derived a median size of $0\farcs30\pm0\farcs04$ ($2.4\pm0.2$~kpc) for the major axis FWHM through Gaussian fits in the $uv$ plane.\footnote{Simpson et al. (2015a) do not tabulate the individual source sizes, and hence we cannot plot the corresponding $\lambda_{\rm obs}=870$~$\mu$m size distribution in our Fig.~\ref{figure:sizes}. We note that Simpson et al. (2015b) list the angular FWHM sizes of these ALMA SMGs in their Table~1, but the source redshifts are not tabulated by Simpson et al. (2015a,b).} The authors pointed out that Gaussian fits in the image plane yielded sizes consistent with those derived in the $uv$ plane, the median ratio between the two being FWHM($uv$)/FWHM(image)$=0.9\pm0.2$. Given the photo-$z$ values of the Simpson et al. (2015a) SMGs, the derived median size refers to that at $\lambda_{\rm rest}\sim250$~$\mu$m. The median angular (linear) size at a comparable rest-frame wavelength from Ikarashi et al. (2015) 
is still $\sim36\%$ ($\sim50\%$) smaller than in the Simpson et al. (2015a) survey. On the other hand, our median 3~GHz angular (linear) major axis FWHM is $1.8\pm0.4$ 
($1.8\pm0.5$) times larger than the median observed-frame 870~$\mu$m size from Simpson et al. (2015a). Similarly, Simpson et al. (2015a) concluded that their 
rest-frame FIR sizes are considerably smaller ($\sim2$ times on average) than the 1.4~GHz radio-continuum sizes from Biggs \& Ivison (2008).

\subsubsection{Size of the CO emission}

In Fig.~\ref{figure:sizes}, we also show the sizes of SMGs as derived through high-resolution ($\sim0\farcs6$) CO spectral line ($J=3-2$ and $7-6$) observations 
with the PdBI by Tacconi et al. (2006). We derived a median CO-emitting FWHM size of $(0\farcs40\pm0\farcs12) \times (0\farcs40\pm0\farcs10)$ or $ (4.1\pm1.0)\times(3.3\pm0.9)$~kpc$^2$ 
for the Tacconi et al. (2006) SMGs that lie at $z_{\rm spec}=2.202-2.509$ and have a reported circular/elliptical Gaussian-fit ($uv$ plane) FWHM size in their Table~1. 
We note that the target SMGs of Tacconi et al. (2006) have SCUBA 850~$\mu$m flux densities of $S_{\rm 850\, \mu m}=8.2-10.7$~mJy, which correspond to 1.1~mm flux densities of 
$S_{\rm 1.1\, mm}\sim3.3-4.3$~mJy (assuming $\beta=1.5$). Hence, in terms of $S_{\rm 1.1\, mm}$, those SMGs are comparable to the faintest sources in our sample (AzTEC21--30) where 
we have only three 3~GHz detections (AzTEC21a, 24b, and 27). Moreover, three of the Tacconi et al. (2006) SMGs appear to be hosting an AGN (SMM J044307+0210, J123549+6215, and J123707+6214). 
Nevertheless, our median 3~GHz major axis FWHM appears to be comparable to the median CO-emission major axis FWHM from Tacconi et al. (2006): the ratio between the two in angular 
and linear units is $1.4\pm0.5$ and $1.0\pm0.3$, respectively. The mid- to high-$J$ CO lines observed by Tacconi et al. (2006) are more sensitive to denser 
and warmer molecular gas than lower excitation ($J_{\rm up}\leq2$) lines, and therefore the total molecular extent is expected to be larger. 
Indeed, one of the Tacconi et al. (2006) SMGs (J123707+6214) was observed in CO$(J=1-0)$ with the VLA by Riechers et al. (2011), 
and it was found to be more spatially extended compared to that seen in CO$(3-2)$ emission. Engel et al. (2010; Table~1 therein) provided a 
compilation of different CO rotational transition observations towards SMGs, and reported linear HWHM sizes for the SMGs as derived 
using circular Gaussian fits in the $uv$ plane (with two exceptions where the quoted size corresponds to the half-light radius). Their target SMGs are characterised by 850~$\mu$m flux densities of $S_{\rm 850\, \mu m}\geq5$~mJy ($S_{\rm 1.1\, mm}\geq2$~mJy if $\beta=1.5$), and this threshold is exceeded by all our SMGs in the JCMT/AzTEC 1.1~mm survey (\cite{scott2008}). From their data we derived a median HWHM value of $1.85\pm0.39$~kpc, which corresponds to a FWHM of $3.70\pm0.78$~kpc, consistent with the aforementioned size we calculated for the Tacconi et al. (2006) SMGs, and hence comparable to our median radio emission size (the median major axis FWHM being $4.2\pm0.9$~kpc).

\subsubsection{The spatial extent of the stellar emission}

Chen et al. (2015) studied the rest-frame optical sizes of SMGs. Based on the \textit{Hubble}/WFC3 observations of a sample of 48 ALMA-detected $z=1-3$ SMGs in the Extended \textit{Chandra} Deep Field South (ALESS SMGs), the authors measured a median effective radius (half-light radius along the semi-major axis within which half of the total flux is emitted) of $4.4^{+1.11}_{-0.5}$~kpc through fitting a S\'ersic profile to the $H_{160}$-band ($\lambda_{\rm pivot}=1\,536.9$~nm) surface brightness of each SMG. Simpson et al. (2015a) compared their FIR sizes to the optical sizes from Chen et al. (2015), and found a large difference of about a factor of four between the two (the optical emission being more extended). Interestingly, the median radius at the rest-frame UV/optical for the AzTEC SMGs from Toft et al. (2014) is only 0.7~kpc (derived using survival analysis; see our Table~\ref{table:othersizes} for the diameters), but we note that most of these sources are very high-redshift SMGs (such as AzTEC1 and AzTEC3), which makes their size determination more difficult, and, as mentioned in Sect.~5.1, the measured sizes are probably subject to strong dust extinction.

\section{Discussion}

\subsection{The spatial extent of SMGs as seen in the radio, dust, gas, and stellar emission}

The radio continuum emission, thermal dust emission, and molecular spectral line emission can all be linked to the stellar evolution process in a galaxy. 
Star formation takes place in molecular clouds where the gas and dust are well mixed. The molecular gas content is best traced by the rotational line emission of CO. 
However, different transitions (arising from different $J$ levels) have different excitation characteristics, hence are probing regions of differing physical and chemical properties: the high-excitation line emission is arising from denser and warmer phase, while low-excitation lines (especially the $J=1-0$ transition) are probing colder, 
more spatially extended gas reservoirs (e.g. \cite{ivison2011}; \cite{riechers2011}). Dust grains absorb the UV/optical photons emitted by the young, newly formed stellar population, and then re-emit the absorbed energy in the FIR. When the high-mass stars undergo SN explosions, the associated blast waves and remnant shocks give rise to synchrotron 
radio emission produced by relativistic CRs. This connection is believed to lead to the tight FIR-radio correlation (see Sect.~1 and references therein). 
On the basis of this connection, one would also expect the FIR- and radio-emission size scales to be similar. The galactic-scale outflows driven by the starburst phenomenon (SNe, 
stellar winds, and radiation pressure) are not expected to overcome the gravitational potential of the galaxy, hence not dispersing the ISM out of the galaxy (this requires a stronger feedback from the AGN; e.g. \cite{tacconi2006}). To summarise, the  radio continuum, rest-frame FIR, and mid- to high-$J$ CO transitions are all expected to trace regions of 
active star formation, and hence the corresponding spatial extents of their emission are expected to be comparable to each other. However, as the size comparison 
in the previous subsection shows, this does not seem to be the case for SMGs. 

As discussed in Sect.~5.2, we have found that the 3~GHz radio-continuum sizes are comparable to the CO-emission sizes from Tacconi et al. (2006) and Engel et al. (2010), 
but more extended than the FIR emission seen in other studies, most notably when compared to those from Ikarashi et al. (2015). A possible scenario is that SMGs have 
a two-component ISM: a spatially extended gas component, which is traced by low- to mid-$J$ CO line emission and radio continuum emission, and a more compact starburst region 
giving rise to the higher-$J$ CO line emission. In the former component, a low dust temperature would lead to a low dust luminosity, while the latter one -- having an elevated 
dust temperature -- could dominate the luminosity-weighted dust continuum size measurements (cf.~\cite{riechers2011}).

Simpson et al. (2015a) suggested that the larger radio-continuum size compared to that at rest-frame FIR is the result of CR diffusion in the galactic magnetic field (e.g. \cite{murphy2008}). To quantify this, they convolved their median 870~$\mu$m size with an exponential kernel and a scale length of 1--2~kpc on the basis of the diffusion length of CR electrons in local star-forming galaxies (which is an order of magnitude longer than the mean free path of dust-heating UV photons; \cite{bicay1990}; \cite{murphy2006}, 2008). The convolved size (FWHM) of 3.8--5.2~kpc they derived is in better agreement with the median major axis FWHM of $6.3\pm0.9$~kpc from Biggs \& Ivison (2008; see our Sect.~5.2). However, as pointed out by Simpson et al. (2015a), the diffusion scale length of CRs in SMGs might be shorter than the aforementioned value because of the higher SFR surface density in SMGs (\cite{murphy2008}; see our Appendix~E). 

The rest-frame FIR sizes of AzTEC1 and AzTEC3 (see Sect.~5.1) suggest that they are comparable to their 3~GHz radio sizes within the uncertainties, but higher-resolution (sub)mm continuum imaging of all our SMGs is required to better constrain their FIR emission sizes, and to examine whether they represent the population of very compact SMGs, similarly to those from Ikarashi et al. (2015). However, even if the radio size is more extended than the FIR emission, the short cooling time of CR electrons in starburst galaxies ($\sim10^4-10^5$~yr) suggests that their diffusion through the ISM to spatial scales larger than FIR emission is infeasible (see Appendix~E for the calculation; the diffusion length ranges from only a few tens of pc to $\sim10^2$~pc). Hence, the CR diffusion scenario proposed by Simpson et al. (2015a) seems unlikely, and in Sect.~6.2 we will discuss a possible alternative explanation for a more extended radio emission in SMGs.  

A further puzzle is the fact that the rest-frame FIR sizes of SMGs appear smaller than the CO-emitting size given the FIR-CO correlation found for different types of galaxies at both low- and high-$z$, including SMGs (see e.g. \cite{carilli2013} for a review; Fig.~7 therein). The large difference found between the rest-frame FIR and optical sizes of SMGs (about a factor of four; see our Sect.~5.2.4) led Simpson et al. (2015a) to conclude that the spatial extent of ongoing star formation is more compact than the spatial distribution of pre-existing stellar population, and that their SMGs might be undergoing a period of bulge growth. As pointed out by Chen et al. (2015), if the high-redshift ($z\gtrsim3$) SMGs are progenitors of $z\sim2$ compact, quiescent galaxies (cQGs; see \cite{toft2014}), the high-$z$ SMGs have to go through a major transformation to decrease the spatial extent of the stellar component (and to increase the S\'ersic index) before being quenched.


\begin{figure}[!h]
\centering
\resizebox{\hsize}{!}{\includegraphics{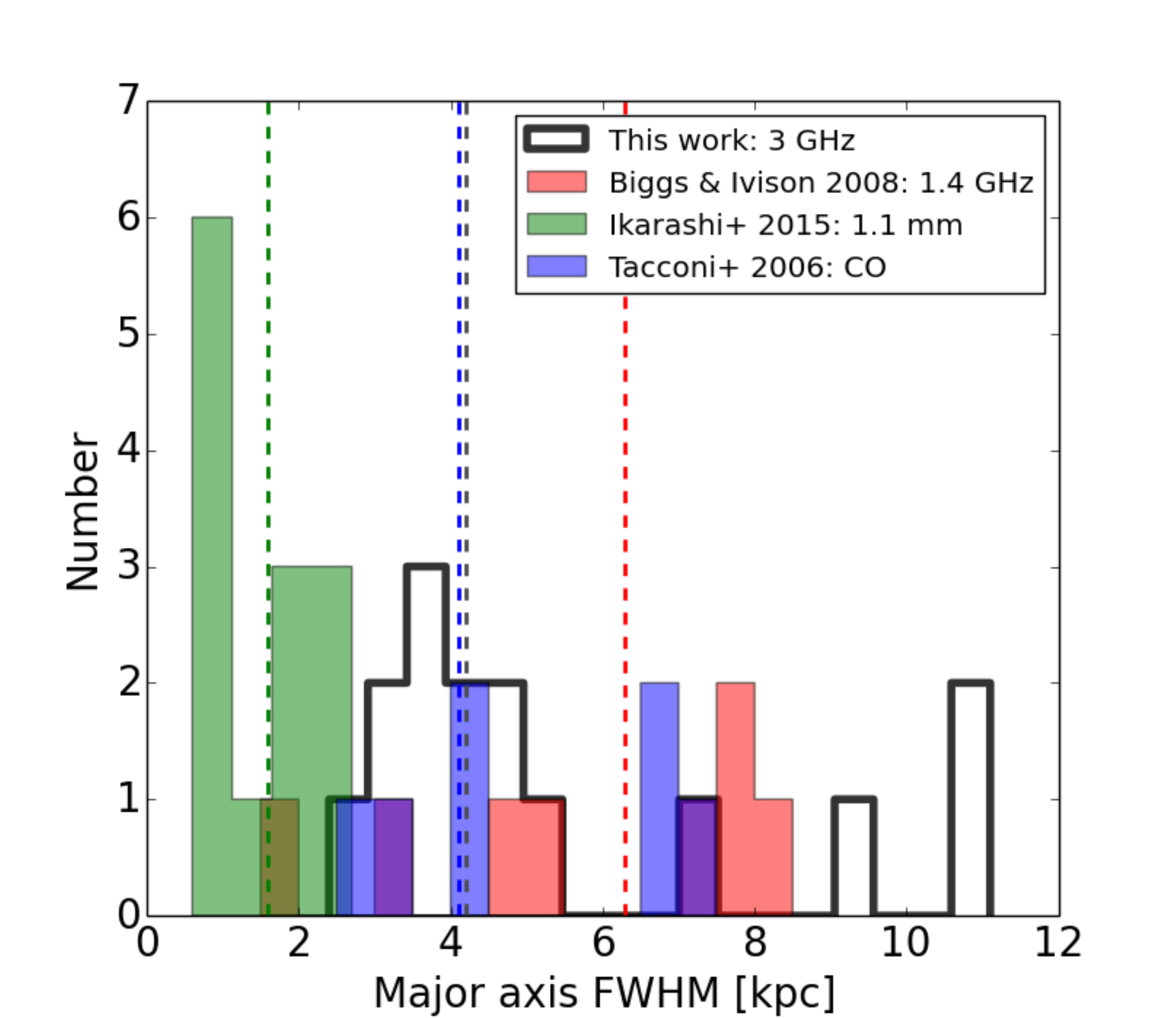}}
\caption{The distribution of the major axis FWHM sizes of our COSMOS SMGs as seen at $\nu_{\rm obs}=3$~GHz is shown by an open histogram. 
For comparison, the following SMG major axis size distributions are also shown: 1.4~GHz sizes from Biggs \& Ivison (2008), 1.1~mm sizes 
from Ikarashi et al. (2015), and the extent of CO molecular gas in the SMGs studied by Tacconi et al. (2006). The upper size limits were placed in
the bins corresponding to those values. The vertical dashed lines show the corresponding median major axis sizes 
[4.2~kpc for AzTEC1--21a, 6.3~kpc for the Biggs \& Ivison (2008) SMGs, 1.6~kpc for the Ikarashi et al. (2015) SMGs, 
and 4.1~kpc for the CO sizes from Tacconi et al. (2006); survival analysis was used to take the upper size limits into account when 
calculating the median sizes]. See text for details.}
\label{figure:sizes}
\end{figure}

\subsection{Merger-induced extended synchrotron emission} 

In the scenario where the FIR size of a galaxy is smaller than its radio-emitting region, and where CR electron diffusion -- due to rapid radiative cooling ($\sim10^4-10^5$~yr) -- is unlikely to be the reason for a more extended radio size (which is potentially the case here; see Appendix~E), an alternative interpretation is required. One possibility is that if SMGs are driven by mergers (e.g. \cite{tacconi2008}; \cite{engel2010}), the interacting progenitor disk galaxies can perturb each others magnetic fields by pulling them out to larger spatial scales (see \cite{murphy2013}). Hence, a significant amount of non-star formation related radio emission can arise from the merger system. Murphy (2013) concluded that this ``taffy''-like merger scenario could explain the low FIR/radio ratios and steep high-frequency radio spectra of local compact starbursts and those seen in some high-$z$ SMGs. In this scenario, mergers are expected to be associated with stretched magnetic field structures 
between the colliding galaxies, giving rise to synchrotron bridges between them and/or tidal tails (\cite{condon1993}). The synchrotron-emitting relativistic electrons 
in such bridges might have their origin in merger-induced shock acceleration, rather than having travelled there from the progenitor galaxies due to the rapid cooling time 
(\cite{lisenfeld2010}; \cite{murphy2013}; see also \cite{donevski2015}). 

The 3~GHz sources investigated here are fairly centrally concentrated and no evidence of interaction-induced bridges/tails is seen except towards AzTEC1, 2, and AzTEC11. 
There is a $\sim2.6\sigma$ 3~GHz feature lying $1\farcs5$ to the NE of AzTEC1, and the 3~GHz major axis FWHM of AzTEC1 
($0\farcs67^{+0.17}_{-0.20}$) is larger than the sample median major axis FWHM ($0\farcs54\pm 0\farcs11)$. 
AzTEC2 exhibits an additional 3~GHz source to the SW, which might be an indication of a merging pair (or a radio jet). The additional source has a major axis FWHM of 
$0\farcs72^{+0.31}_{-0.44}$, which is also larger than the median $\theta_{\rm maj}$ of $0\farcs54\pm 0\farcs11$. The two 3~GHz components seen towards AzTEC11 share a common 
$3\sigma$ 3~GHz envelope, but AzTEC11-N and 11-S both have a 3~GHz major axis FWHM size smaller than the median value ($0\farcs40^{+0.07}_{-0.08}$ and $0\farcs48^{+0.04}_{-0.05}$). The 1.4~GHz morphologies of the Biggs \& Ivison (2008; their Fig.~3) SMGs are generally more elongated and clumpy than our sources, which could suggest a higher merger fraction among their SMGs, and hence somewhat more extended radio emission sizes (see Sect.~5.2.1). 
However, a fair fraction of our target SMGs ($\sim36\%$ of the total sample) show clumpy/disturbed morphologies or evidence of close companions at different 
wavelengths (\cite{younger2007}, 2009; \cite{toft2014}; \cite{miettinen2015}), which could be manifestations of galaxy mergers.


To conclude, there could be a possible connection between merger-driven SMGs and their larger radio-emitting size as compared to FIR emission, 
as would be expected if the above described merger scenario is true. However, the spatial distribution of molecular gas, as traced by mid- to high-$J$ lines, 
appears to be comparable to the $\nu_{\rm obs}=3$~GHz radio emission size. As described in Sect.~6.1, this is to be expected if the observed radio size of a galaxy is a direct tracer of its spatial extent of star formation. This would not be consistent with the scenario where the CRs emit synchrotron radiation as a result of processes \textit{not} related to star formation, such as the aforementioned merger scenario. However, these comparisons between CO and radio median sizes are, unfortunately, based on measurements obtained from different samples and the result can be affected by subtle selection effects. For example, the 1.4~GHz radio sizes from Biggs \& Ivison (2008) are instead larger than the CO sizes from Tacconi et al. (2006) (see our Fig.~\ref{figure:sizes}), which is qualitatively consistent with the scenario of merger-induced synchrotron emission. To quantitatively compare the spatial extents of radio emission and molecular gas component, high-resolution radio and CO imaging of the \textit{same} sample of SMGs is required.


\subsection{Size evolution as a function of redshift and the effect of galaxy environment}

\begin{figure}[!h]
\centering
\resizebox{\hsize}{!}{\includegraphics{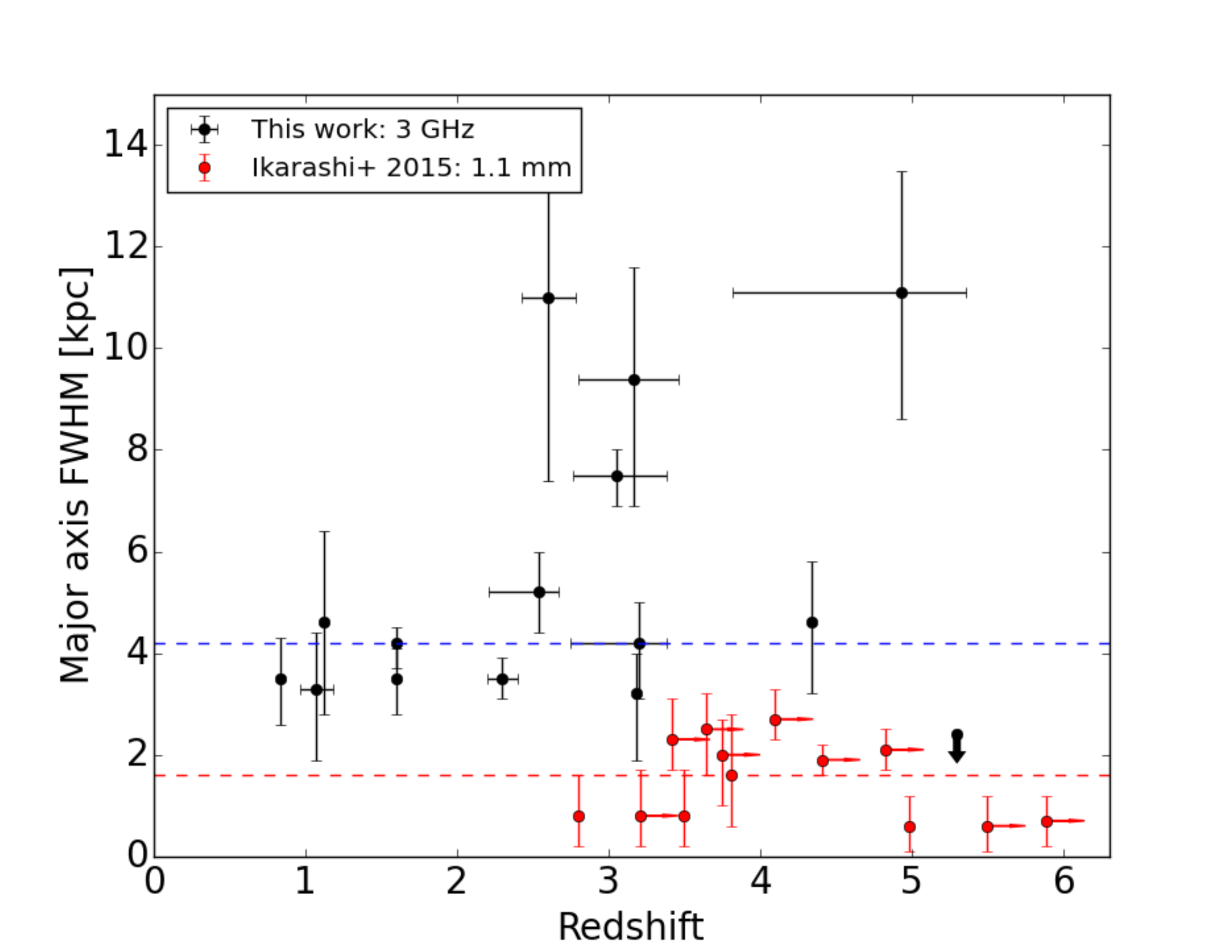}}
\caption{The linear major axis FWHM sizes [kpc] of our AzTEC SMGs at 3~GHz plotted as a function of redshift. 
The upper size limit of AzTEC3 is indicated by a downward pointing arrow. Also shown (red points) are the 1.1~mm sizes 
from Ikarashi et al. (2015). The horizontal blue and red dashed lines show the corresponding median major axis FWHM values of 4.2~kpc and 1.6~kpc, respectively.}
\label{figure:corr}
\end{figure}

In Fig.~\ref{figure:corr}, we show our deconvolved linear major axis FWHM sizes as a function of redshift. 
No statistically significant correlation can be seen between these two quantities, which is consistent with that found by Simpson et al. (2015a) 
and Chen et al. (2015) at shorter wavelengths. We note, however, that, with the exception of AzTEC3 (see below), 
there is a hint of larger radio sizes at $z\sim2.5-5$ compared to our lower redshift SMGs: the $z\sim2.5-5$ SMGs tend to lie above the median size of our sample 
(4.2~kpc, blue dashed line). 
 Also plotted in Fig.~\ref{figure:corr} are the 1.1~mm FWHM sizes from Ikarashi et al. (2015). These authors discussed that the compact sizes of 
their high-redshift ($z_{\rm phot}\sim 3-6$) SMGs support the scenario where they represent the precursors of cQGs seen at $z\sim 2$, which, in turn, are believed to 
evolve into the massive ellipticals seen in the present-day ($z=0$) universe (\cite{toft2014}). We note that among the Ikarashi et al. (2015) SMG sample, 
the $\lambda_{\rm obs}=1.1$~mm sizes are larger at $z\sim3.5-5$ than those outside that redshift range, although it should be noted that most of their SMGs 
in this $z$ range have only lower $z$ limits available. As mentioned above, there is some resemblance in our data, i.e. the radio sizes appear larger 
at a comparable redshift range of $z\sim2.5-5$.

Ikarashi et al. (2015) discussed that if both the radio and FIR continuum are tracers of star-forming regions, then the $z\gtrsim 3$ SMGs are more compact than the lower-redshift SMGs typically observed in radio continuum emission (e.g. \cite{biggs2008}). As shown in Fig.~\ref{figure:corr}, our present VLA 3 GHz data do not suggest such a trend, and, as mentioned earlier, there is actually a hint of larger radio sizes at $z\sim2.5-5$ compared to lower redshifts. However, the highest-redshift SMG in our sample, AzTEC3 at $z\simeq5.3$, shows the most compact size among our sources, consistent with the rest-frame FIR sizes from Ikarashi et al. (2015). We note that Capak et al. (2011) found that AzTEC3 belongs to a spectroscopically confirmed protocluster containing eight galaxies within a 1~arcmin$^2$ area, and therefore the environment might also play a role in the galaxy size evolution (see also \cite{smolcic2015b}). However, it is currently unclear whether the environmental effects in a galaxy overdensity will lead to a more compact \textit{or} more extented radio-emitting size compared to field galaxies. On one side, a protocluster environment is expected to show an elevated merger rate (e.g. \cite{hine2015}), and, as discussed above, mergers are expected to pull the galactic magnetic fields to larger spatial scales, and hence lead to a more extended radio synchrotron emission. On the other side, the ram and/or thermal pressures of the intracluster medium could compress the ISM of the galaxy, increase the magnetic field strength, and hence cause an excess in radio emission (consistent with a low IR-radio $q$ parameter of $\lesssim2$ for AzTEC3; O.~Miettinen et al., in prep.). The aforementioned pressure forces can drive shock waves into the ISM, and hence accelerate the CR particles (\cite{murphy2009}). Consequently, the cooling time and diffusion length-scale of CR electrons can decrease (see Appendix~E), resulting in a compact radio-emitting area. More detailed environmental analysis of SMGs is needed to understand this further. 

\section{Summary and conclusions}

We have used radio-continuum observations taken by the VLA-COSMOS 3~GHz Large Project to study the radio sizes of 
a sample of SMGs originally detected with the AzTEC bolometer array at 1.1~mm, and all followed up with (sub)mm interferometric 
observations. Our main results are summarised as follows:

\begin{enumerate}
\item Of the total sample of 39 SMGs, 3~GHz emission was detected towards 18 or $\sim46\pm11\%$ of them (${\rm S/N=4.2-37.4}$). Four sources (AzTEC2, 5, 8, and 11) 
show two separate 3~GHz sources.
\item The median angular radio-emitting size (FWHM) we derived is $(0\farcs54\pm 0\farcs11)\times(0\farcs35\pm0\farcs04)$. In linear units, derived for the SMGs 
with known redshift, we obtained a median size of $(4.2\pm0.9)\times(2.3\pm0.4)$~kpc$^2$. 
 The low brightness temperature values of $T_{\rm B}=1.4\pm0.7$~K to $75.9\pm16.6$~K are consistent with the radio emission being powered by star formation, rather than by an AGN. 
\item We found no obvious correlations between the FWHM radio-size and radio or submm flux density or redshift, which is consistent with previous studies at other wavelengths. 
\item We found that our derived 3~GHz sizes are comparable to 1.4~GHz and CO-emission sizes of SMGs reported in literature, yet they are $\sim1.7-2.8$ times larger than the median rest-frame FIR sizes based on high-resolution ALMA observations, and reported in literature (see Sect.~5.2 for details).
\item If both the radio and FIR continuum are tracing the same regions of star formation in a galaxy as expected from the FIR-radio correlation, 
then the differing spatial scales of these emissions is puzzling. A possible explanation is that SMGs have a two-component ISM: \textit{i)} an extended gas component with a low dust temperature, which gives rise to the low- to mid-$J$ CO line and radio continuum emissions, and \textit{ii)} a warmer, compact starburst region giving rise to the high-$J$ rotational line emission of CO, which could dominate the dust continuum size measurements.
The more extended radio-emitting size with respect to the compact FIR-emitting size was suggested to be the result of cosmic-ray diffusion by Simpson et al. (2015a). However, we have shown here that the short electron cooling times of $\sim10^4-10^5$~yr in dense starburst galaxies do not allow the electrons to spread away from their sites of origin to the required spatial scales. Hence, it seems more probable that the observed synchrotron emission partly originates in regions around the active starburst region, possibly from extended magnetic fields driven by the galaxy merging process. 
\end{enumerate}


\begin{acknowledgements}

We thank the referee for constructive comments that helped to improve this paper. 
This research was funded by the European Union's Seventh Framework programme 
under grant agreement 337595 (ERC Starting Grant, 'CoSMass'). 
M.~A. acknowledges partial support from FONDECYT through grant 1140099. 
AK acknowledges support by the Collaborative Research Council 956,
sub-project A1, funded by the Deutsche Forschungsgemeinschaft (DFG). 
This paper makes use of the following ALMA data: ADS/JAO.ALMA\#2012.1.00978.S and ADS/JAO.ALMA\#2013.1.00118.S. 
ALMA is a partnership of ESO (representing its member states), NSF (USA) and NINS (Japan), together with NRC (Canada), 
NSC and ASIAA (Taiwan), and KASI (Republic of Korea), in cooperation with the Republic of Chile. 
The Joint ALMA Observatory is operated by ESO, AUI/NRAO and NAOJ. 
This research has made use of NASA's Astrophysics Data System, and the NASA/IPAC Infrared Science Archive, 
which is operated by the JPL, California Institute of Technology, under contract with the NASA. 
We greatfully acknowledge the contributions of the entire COSMOS collaboration consisting of more than 100 scientists. 
More information on the COSMOS survey is available at {\tt http://www.astro.caltech.edu/$\sim$cosmos}. 

\end{acknowledgements}

\appendix

\section{Notes on peculiar 3~GHz sources}

AzTEC2, 5, and 8 each show, in addition to a 3~GHz source coinciding with the submm peak, an additional source 
lying at $1\farcs51$ SW, $6\farcs56$ NE, and $2\farcs62$ NE from the SMA peak, respectively. 
The additional 3~GHz feature towards AzTEC2 could, in principle, represent a radio-emitting lobe of a jet interacting with 
the surrounding medium or a merger component (projected separation is 12.7 proper kpc at the redshift of AzTEC2). 
We note that the 1.3~mm emission detected towards AzTEC2 with ALMA also shows an additional weak ($\sim2.9\sigma$) feature at 
$2\farcs14$ to the SW of AzTEC2 (Cycle~2 ALMA project 2013.1.00118.S; M.~Aravena et al., in prep.), but its peak position 
lies $0\farcs60$ away to the SW of the 3~GHz feature; this offset is within the large statistical positional 
uncertainty of $\sim1\arcsec$ of the weak 1.3~mm feature [$\Delta \theta_{\rm stat} \propto ({\rm S/N})^{-1}$]. 

The additional radio source towards AzTEC5 was already seen at 1.4~GHz (see Table~\ref{table:radio}), 
and it is also visible in the \textit{Spitzer}/IRAC and MIPS images 
(\cite{younger2007}, Fig.~1 therein). This source can be associated with the \textit{Herschel}-detected 
emission-line galaxy 150.08336+02.53619, for which a spectroscopic redshift of $z_{\rm spec}({\rm H\alpha})=1.42$ was reported by Roseboom et al. (2012). 
Also, a $5.4\sigma$ detection with ALMA at 994~$\mu$m is obtained towards this source (Cycle~1 ALMA project 2012.1.00978.S; PI: A.~Karim; O.~Miettinen et al., in prep.), 
while the ALMA 1.3~mm detection is of $\sim3\sigma$ significance (M.~Aravena et al., in prep.). We note that the most up-to-date COSMOS spec-$z$ catalogue gives 
a lower redshift value of $z_{\rm spec}=0.9044$, which is based on observations with the Inamori Magellan Areal Camera 
and Spectrograph (IMACS; M.~Salvato et al., in prep.); however, the quality flag is 1, i.e. this $z_{\rm spec}$ is considered insecure.
Hence, the 3~GHz source NE of AzTEC5 is probably a lower-redshift galaxy (AzTEC5 itself has $z_{\rm phot}=3.05_{-0.28}^{+0.33}$; 
see Table~\ref{table:sample}). Similarly, the two radio sources towards AzTEC8 were already seen at 1.4~GHz (\cite{younger2009}, Fig.~1 therein; see our Table~\ref{table:radio}):
the 1.4~GHz source to the NE of the SMA-detected SMG was called AzTEC8.E by Younger et al. (2009), and the \textit{Spitzer} 24~$\mu$m emission towards AzTEC8 is coincident with AzTEC8.E. 
This source can be associated with the \textit{Chandra}/X-ray -detected galaxy CXOC J095959.5+023441 whose photo-$z$ is $z_{\rm phot}=2.420\pm0.060$ (\cite{salvato2011}), 
hence it is probably at a lower redshift than the SMG on its western side ($z_{\rm spec}=3.179$). This source is also detected with the VLBA at 1.4~GHz 
at milliarcsec resolution ($S_{\rm 1.4\, GHz}=83.8$~$\mu$Jy), showing the presence of a radio-emitting AGN (N.~Herrera Ruiz et al., in prep.; N.~Herrera Ruiz, priv.~comm.).

Towards AzTEC11, we have detected a double 3~GHz source, projectively separated by $1\farcs5$, and where the southern component is coincident with 
AzTEC11-N\footnote{We note that the southern SMA source towards AzTEC11 was accidentally called AzTEC11.N by Younger et al. (2009), while the northern component 
was called AzTEC11.S (see Table~1 in \cite{younger2009}). }. The northern 3~GHz source is nearly equidistant 
from AzTEC11-N ($1\farcs38$ separation) and AzTEC11-S ($1\farcs40$ separation). Younger et al. (2009) reported that the calibrated visibility data of AzTEC11 show 
significant structure and are best modelled with a double point source (their Table~2). The complex structure of the visibility data probably makes the derived 
source positions to be rather uncertain. However, a positional uncertainty of only $0\farcs2$ in both right ascension and declination for the 890~$\mu$m peak of AzTEC11-N 
and 11-S was reported by Younger et al. (2009; Table~1 therein). The authors also recognised an elongated 1.4~GHz source towards AzTEC11, where the emission morphology resembles 
that seen at 890 $\mu$m. Their two-component Gaussian fit yielded comparable 1.4~GHz flux densities for the two sources (see our Table~\ref{table:radio}). 
The two 3~GHz sources seen towards AzTEC11 share a common radio envelope (at the $3\sigma$ level), and are probably in the process of merging\footnote{The spectroscopic redshift of 
$z_{\rm spec}=1.599$ was measured towards a position, which lies $0\farcs6$ NE of AzTEC11-N's 3~GHz peak position, and $0\farcs9$ SW from that of AzTEC11-S 
(M.~Salvato et al., in prep.). The physical relation of the 3~GHz sources is supported by the very low probability for a chance association. This can be quantified by calculating the Poissonian random probability, $P=1-e^{-\pi r^2 N}$, where $r$ is the projected angular distance of the two sources, and $N$ is the surface number density of sources [deg$^{-2}$] that have flux densities greater than or equal to that of the source in question (\cite{downes1986}; \cite{scott1989}). Using the VLA-COSMOS 3 GHz Large Project catalogue (V.~Smol{\v c}i{\'c} et al., in prep.), we estimate that the probability of having a AzTEC11-S type 3~GHz source lying $1\farcs5$ away from that of AzTEC11-N is only $P\sim3.4\times10^{-4}$. The null hypothesis of chance association is generally rejected if $P<5\%$.}. The ALMA 1.3~mm observations at $1\farcs36 \times 0\farcs78$ resolution towards AzTEC11 revealed two SMGs separated by $1\farcs45$ in projection 
(M.~Aravena et al., in prep.), and the northern one is well coincident with our northern 3~GHz source (only $0\farcs11$ offset).

Towards AzTEC15 the projected separation between the 890~$\mu$m and 3~GHz positions is fairly large, i.e. $1\farcs19$. The 890~$\mu$m detection of AzTEC15 by Younger et al. (2009) was 
only of moderate significance ($4.4\sigma$), and no 1.4~GHz counterpart was detected, but it was found to be associated with \textit{Spitzer} IR emission. 
The 890~$\mu$m position uncertainty reported by Younger et al. (2009) is $0\farcs3$ in right ascension and $0\farcs2$ in declination. The $1\farcs36 \times 0\farcs78$ resolution 
ALMA 1.3~mm observations towards AzTEC15 (M.~Aravena et al., in prep.), however, show a perfect positional coincidence ($0\farcs03$ offset) with our 3~GHz source of $5.4\sigma$ 
significance, and hence it is physically related to AzTEC15.

The 3~GHz source near AzTEC21a ($1\farcs15$ NE of the PdBI position) is our weakest source with a S/N ratio 
of 4.2. The reliability of this 3~GHz source candidate is supported by the fact that it is also seen at 1.4~GHz, although the 1.4~GHz source is also weak 
(peak surface brightness of 63~$\mu$Jy~beam$^{-1}$ or $\sim3.9\sigma$; see \cite{miettinen2015}, Fig.~A.1 therein). Hence, we include the 3~GHz source near AzTEC21a in 
our radio size analysis.

The 3~GHz source ($12.6\sigma$) detected $2\farcs67$ SW in projection from AzTEC24b is 
also detected at 1.4~GHz (Table~\ref{table:radio}), and the 1.4~GHz source was associated with the ASTE/AzTEC 1.1~mm SMG AzTEC/C48 by Aretxaga et al. (2011), 
a source also detected with \textit{Herschel} (see \cite{miettinen2015}). There is also a 1.3~mm-detected ALMA source lying $2\farcs63$ SW from our 
PdBI detection and $0\farcs26$ SE from the 3~GHz source in question (M.~Aravena et al., in prep.). The ALMA detection in particular shows that the 3~GHz source is an SMG despite the relatively large separation from our PdBI source. We note that there is a $4.3\sigma$ 3~GHz source lying $0\farcs47$ N of AzTEC27. 
Miettinen et al. (2015) reported the presence of weak 1.4~GHz emission (peak intensity of 32.1~$\mu$Jy~beam$^{-1}$ or $\sim2.5\sigma$) towards this position, but the nature of this radio 
emission remains unclear (i.e. noise feature or associated with the SMG). The weak 3~GHz source associated with AzTEC27 is included in the present radio size analysis.
The additional 3~GHz radio sources not analysed further in the present study are described in Appendix~B, and those not detected at 3~GHz are discussed in Appendix~C.

\section{Notes on additional 3~GHz sources}

As shown in Fig.~\ref{figure:maps}, $5\farcs70$ SW from AzTEC20, $8\farcs43$ NW from AzTEC22, 
and $5\farcs51$ E from AzTEC23, there is a clearly detected 3~GHz source ($10.3\sigma$, 
$22.8\sigma$, and $5.5\sigma$, respectively). Interestingly, these 3~GHz sources are closer to the original 
AzTEC 1.1~mm positions than to the PdBI source candidates ($0\farcs75$ E, $2\farcs73$ SW, 
and $4\farcs25$ NE from AzTEC20, 22, and 23; see \cite{scott2008}; \cite{miettinen2015}). Morever, the 3~GHz sources detected towards 
the AzTEC20, 22, and 23 fields have a \textit{Herschel} 250~$\mu$m detection lying at $2\farcs45$ SW, $1\farcs48$ SE, and $7\farcs78$ NW, respectively 
(as based on the cross-correlation with the COSMOS SPIRE 250~$\mu$m Photometry Catalogue). These radio sources are not analysed further in the present study.

\section{Notes on the 3~GHz non-detections' appearances at other wavelengths}

The following 21 SMGs ($54\pm12\%$ of the whole sample) were not detected at 3~GHz: 
AzTEC10, 13, 14-E, 14-W, 16, 17b, 18, 19b, 20, 21b, 21c, 22, 23, 24a, 24c, 26a, 26b, 28, 29a, 29b, and 30. 

AzTEC10, 13, 14-E, and 14-W were detected at 890~$\mu$m with a S/N ratio of 5.3, 4.6, 5.0, and 3.9, respectively, 
by Younger et al. (2009). Only the most significant of these moderate SMA 890~$\mu$m detections, 
namely AzTEC10, was found to exhibit \textit{Spitzer} IR emission, while none of them was detected 
at optical wavelengths or at 1.4~GHz (\cite{younger2009}). Hence, their non-detection at 3~GHz is not surprising. 

As discussed by Miettinen et al. (2015; Appendix~C therein), the PdBI 1.3~mm SMG candidates 
AzTEC16, 17b, 20, 21c, 22, 24a, 24c, 26b, 29a, and 30 have no multiwavelength counterparts and some of them could be spurious. 
The 1.3~mm S/N ratios of these sources were found to be in the range of 4.5--6. 
Moreover, AzTEC29b, although a $7.3\sigma$ detection, was found to lie at edge of the 1.3~mm map. 
On the other hand, AzTEC27 and AzTEC28 (${\rm S/N}_{\rm 1.3\, mm}=6$ and 5.5, respectively) were among 
the best PdBI detections by Miettinen et al. (2015), but neither of them were found to have multiwavelength 
counterparts; only a trace of 1.4~GHz emission ($2.5\sigma$) was seen towards AzTEC27 (Appendix~A). 
AzTEC18, 19b, 21b, 21c, 23, and 26a were detected with ${\rm S/N}_{\rm 1.3\, mm}=4.2-9.7$, 
and only AzTEC21c was found to \textit{not} have multiwavelength counterparts. However, none of these SMGs 
was detected at 1.4~GHz. Given the aforementioned properties, it comes as no surprise that among AzTEC16--30 there 
are so many 3~GHz non-detections (17 of 22, i.e. $77\pm19\%$).

\section{Testing the reliability of the size measurements} 

\begin{figure}[!h]
\centering
\resizebox{0.8\hsize}{!}{\includegraphics{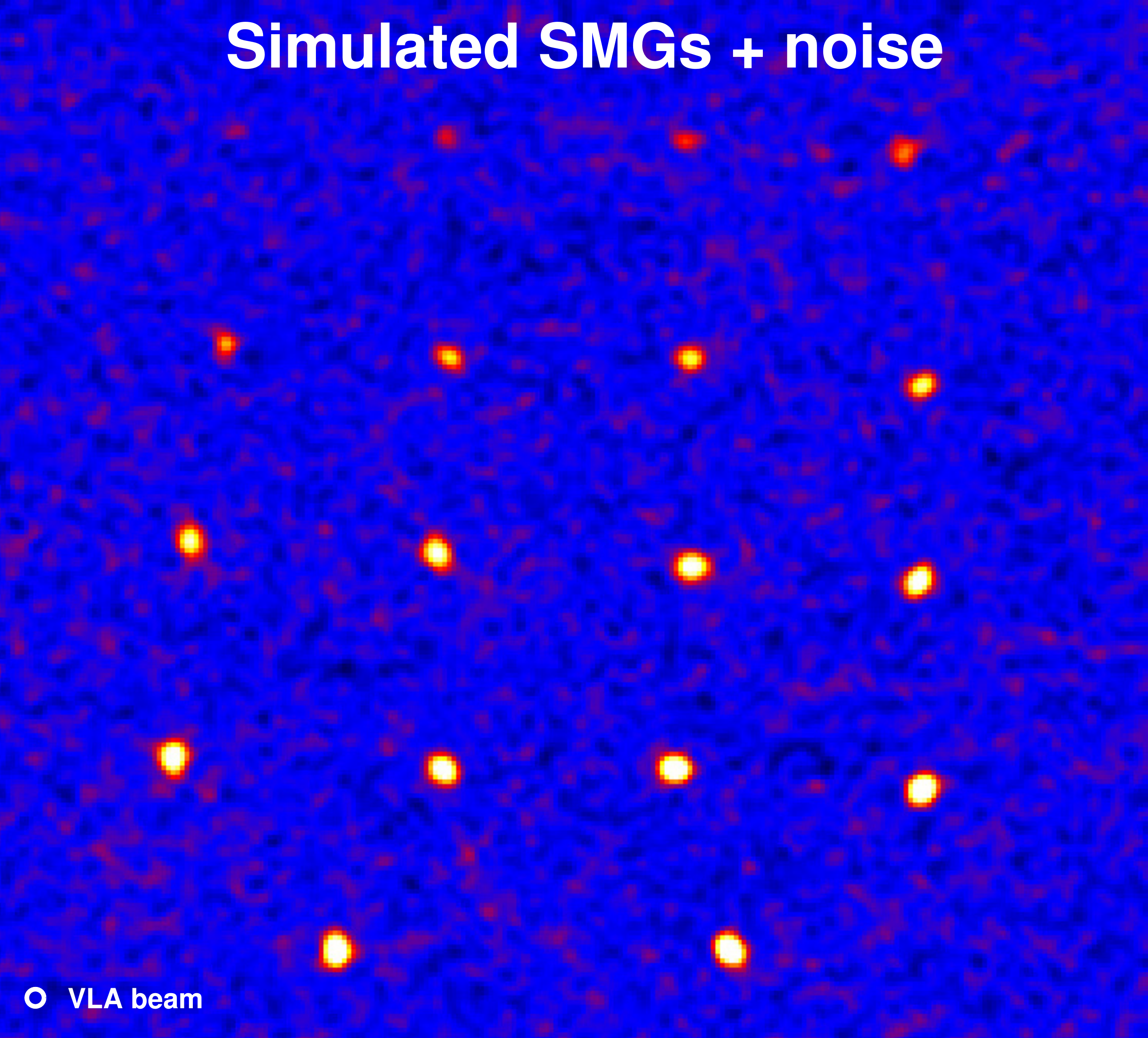}}
\resizebox{\hsize}{!}{\includegraphics{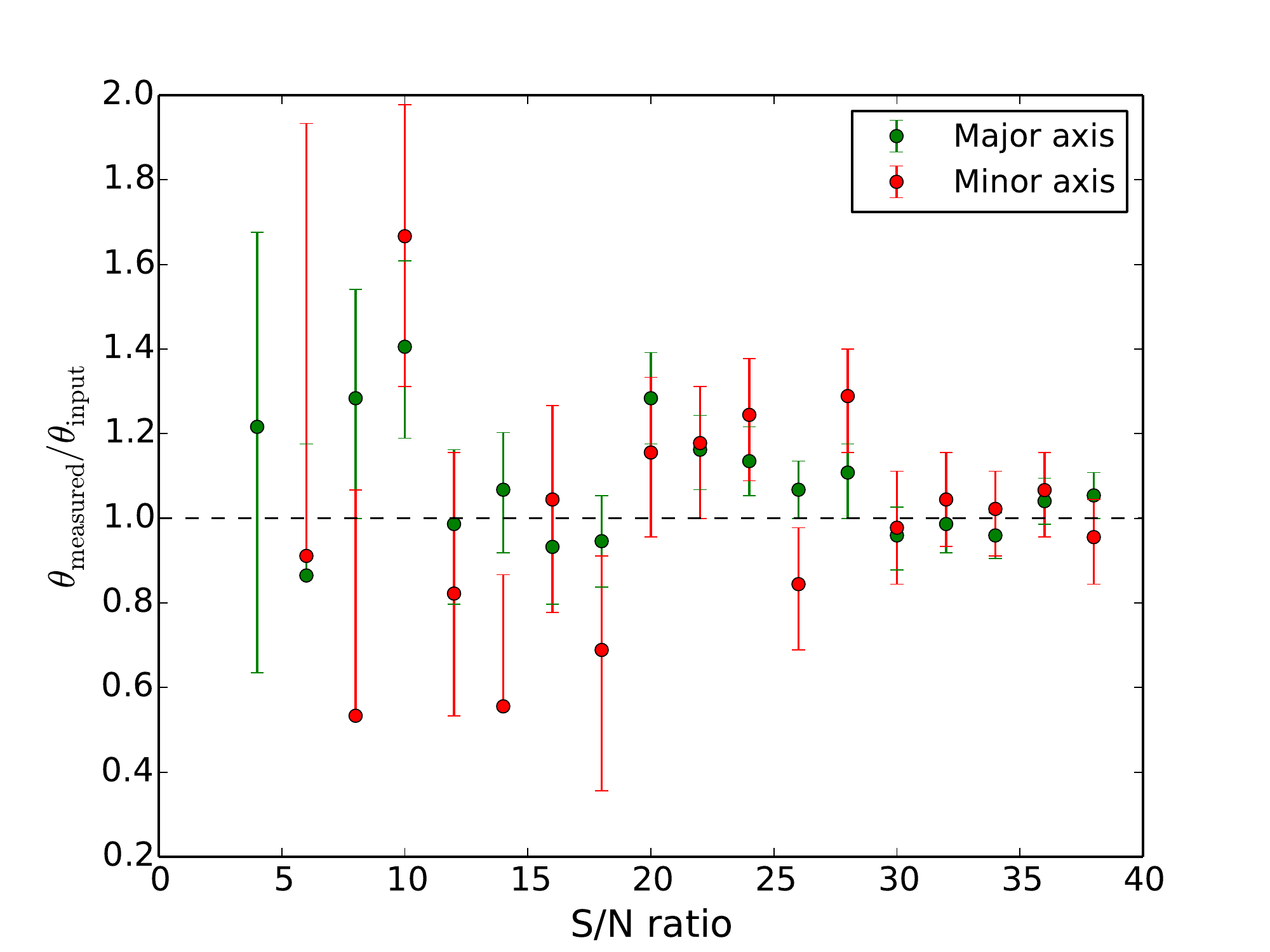}}
\caption{\textbf{Top:} Simulated SMGs added to a noise map to simulate our real 3~GHz VLA data. The S/N ratio of the sources increases 
from left to right, top to bottom (${\rm S/N}=4-38$), and in each row the P.A. ranges from $0\degr$ to $135\degr$ in steps of $45\degr$ 
(being $0\degr$ and $45\degr$ for the two bottommost objects). The synthesised beam size of $0\farcs75$ is shown in the bottom left corner. 
\textbf{Bottom:} The ratio of the measured to the simulated input source size (deconvolved FWHM) as a function of the S/N ratio. 
The green points show the major axes ratio, while the red points show that between the minor axes. The minor axis FWHM for the faintest source 
(${\rm S/N}=4$) could not be determined by the AIPS task {\tt JMFIT}. The vertical error bars were propagated from the size errors determined by 
{\tt JMFIT}. A one-sided error bar is shown for those cases where the minimum size could not be determined by {\tt JMFIT}. 
The horizontal dashed line shows the line of equality between the sizes.}
\label{figure:sim}
\end{figure}

To test the reliability of our FWHM size measurements, we simulated sources using the {\tt CASA} (release 4.3.1) Toolkit. 
We first generated mock galaxies with a Gaussian flux distribution, intrinsic FWHM size fixed to 
$0\farcs74 \times 0\farcs45$ (i.e. the average deconvolved FWHM derived for our SMGs where both the major and 
minor axes could be determined or constrained), P.A. ranging from $0\degr$ to $135\degr$ in steps of $45\degr$, 
and flux densities corresponding to S/N ratios in the range of 4 to 38 in steps of ${\rm S/N}=2$, 
which cover the observed range of S/N ratios of our SMGs (${\rm S/N}=4.2-37.4$). 
The simulated image was convolved to the resolution of $0\farcs75$ 
to match the resolution of our real data. To obtain a realistic background noise level, 
the simulated galaxies were added to a 3~GHz map of $1\farcm5 \times 1\farcm5$ in size, 
and which was cropped from a source-free region of COSMOS ($1\sigma=2.3$~$\mu$Jy~beam$^{-1}$). 
The resulting image is shown in the top panel in Fig.~\ref{figure:sim}. The deconvolved FWHM sizes 
of the simulated sources were then determined using the AIPS task {\tt JMFIT} as described in Sect.~4.1. 

The bottom panel in Fig.~\ref{figure:sim} shows the ratio of the measured size to the input size 
as a function of the S/N ratio. The data points are shown separately for the 
major and minor axes. As expected, the size measurement is generally more accurate for more significant detections, 
but within the size uncertainties determined by {\tt JMFIT} the measured deconvolved sizes are in good agreement 
with the real intrinsic sizes (see the dashed line in Fig.~\ref{figure:sim} 
indicating the one-to-one correspondence). 
Because most of our detections are of high significance (median ${\rm S/N}=12.6$), 
these simulations suggest that our size measurements are reliable.

\section{Calculation of the cosmic-ray electron cooling times} 

To quantitatively examine the possibility that the observed radio-continuum sizes of SMGs could be the result of 
CR diffusion, we first calculate the maximum lifetime of the electrons considering the radiative energy losses due to synchrotron emission, 
inverse Compton (IC) scattering, bremsstrahlung, and ionisation processes. In what follows, we calculate the corresponding cooling times using the formulas 
from Murphy (2009; Sect.~2 therein) to which we refer the reader for a more detailed description. 

The redshifts of our 3~GHz detected SMGs range from $z_{\rm spec}=0.834$ for AzTEC17a to $z_{\rm spec}=5.298$ for AzTEC3. If we assume that 
the critical frequency at which the electron emits most of its energy, $\nu_{\rm crit}$, is equal to $\nu_{\rm rest}=\nu_{\rm obs}(1+z)$, its value is in the range of 
5.5~GHz to 18.9~GHz. If the electrons are spiralling in a magnetic field whose strength is $B\sim100$~$\mu$G (a starburst-type $B$-field; 
e.g. \cite{lacki2013}), the relationship $\nu_{\rm crit}=1.3\times10^{-2}(B/{\rm \mu G})(E/{\rm GeV})^2$ yields CR electron energies of 
$E\simeq2-3.8$~GeV. For this case, the synchrotron cooling timescale for CR electrons is 
$\tau_{\rm syn}\sim1.4\times10^{9}(\nu_{\rm crit}/{\rm GHz})^{-1/2}(B/{\rm \mu G})^{-3/2}\sim3.2-6\times10^5$~yr. 

The IC cooling timescale is given by $\tau_{\rm IC}\sim5.7\times10^{7}(\nu_{\rm crit}/{\rm GHz})^{-1/2}(B/{\rm \mu G})^{1/2}(U_{\rm rad}/{\rm 10^{-12}~erg~cm^{-3}})^{-1}$~yr, where 
$U_{\rm rad}$ is the radiation field energy density of the galaxy. To estimate the value of $U_{\rm rad}$, we adopt a total infrared (8--1\,000~$\mu$m) 
luminosity range of $L_{\rm IR} \sim 10^{12}-10^{13}$~L$_{\sun}$ appropriate for SMGs (e.g. \cite{magnelli2012}; \cite{swinbank2014}; \cite{dacunha2015}; see O.~Miettinen et al., in prep., for the present SMG sample), and as the characteristic size we use the median 3~GHz major axis FWHM size derived here, i.e. 4.2~kpc, which corresponds to a circular area of 13.9~kpc$^2$. Using Eq.~(4) of Murphy et al. (2012a), these values imply $U_{\rm rad}$ in the range of 
$\sim 5 \times 10^{-10}- 4.9 \times 10^{-9}$~erg~cm$^{-3}$; for a smaller IR-emitting area ($A_{\rm IR}$), the value of $U_{\rm rad} \propto A_{\rm IR}^{-1}$ would 
be higher. The values of $\nu_{\rm crit}$ and $B$ being as above, we derive $\tau_{\rm IC}\sim  2.6\times 10^4- 4.9\times 10^5$~yr. In the context of IC cooling, it should be noted that our SMG sample contains high-redshift sources, the most extreme case being AzTEC3 at $z_{\rm spec}=5.298$. 
At high redshifts, the IC scattering between relativistic electrons and the cosmic microwave background (CMB) -- boosting the CMB photon energy -- becomes more 
important compared to the low-$z$ universe. The reason for this is that the energy density of the CMB increases steeply with redshift, namely $U_{\rm CMB}\propto(1+z)^4$. 
For instance, at $z_{\rm spec}=5.298$, the CMB photon bath has about 140 times higher energy density 
than that at the lowest-redshift SMG in our sample (AzTEC17a at $z_{\rm spec}=0.834$). This means that besides the intense IR radiation field in a starburst, 
IC losses off the CMB photons have the potential to increase the cooling rate of the CR electrons (e.g. \cite{lacki2010}). 

The bremsstrahlung lifetime is $\tau_{\rm brem}\sim8.6\times10^{7}(n_{\rm H}/{\rm cm}^{-3})^{-1}$~yr, where $n_{\rm H}$ is the hydrogen number density of the ISM. 
Assuming a typical average density range of $n_{\rm H}=10^2-10^3$~cm$^{-3}$, we obtain $\tau_{\rm brem}\sim 8.6\times10^4-8.6\times10^5$~yr.\footnote{We note that under the assumption of a magnetic flux freezing, i.e. $B \propto n_{\rm H}^{1/2}$ (e.g. \cite{crutcher1999}), our adopted field strength of 100~$\mu$G would imply a density of $n_{\rm H}\simeq317$~cm$^{-3}$ if $B=10$~$\mu$G at $n_{\rm H}=1$~cm$^{-3}$ as is typically observed in normal star-forming galaxies (cf.~\cite{murphy2009}). Our adopted range of densities brackets this value of $n_{\rm H}$.} 

The timescale for ionisation losses can be written as 
$\tau_{\rm ion}\sim3.6\times10^{10}(\nu_{\rm crit}/{\rm GHz})^{1/2}(B/{\rm \mu G})^{-1/2}(n_{\rm H}/{\rm cm}^{-3})^{-1}\times \left[3/2\times \ln (\nu_{\rm crit}/B) +49 \right ]^{-1}$~yr, which under 
our assumptions lies in the range of $\tau_{\rm ion} \sim1.9\times10^5-3.4\times10^6$~yr. 

Finally, due to the combined energy losses from the aforementioned processes, the effective cooling lifetime for CR electrons is given by 

\begin{equation} 
\tau_{\rm cool}=\frac{1}{\tau_{\rm syn}^{-1}+\tau_{\rm IC}^{-1}+\tau_{\rm brem}^{-1}+\tau_{\rm ion}^{-1}}\,.
\end{equation} 
The individual timescales calculated above yield $\tau_{\rm cool}\sim  1.7\times10^4- 1.9\times10^5$~yr. In the case of random-walk diffusion, 
the electrons' escape scale-length is given by $l_{\rm esc}=(D_{\rm E}\tau_{\rm esc})^{1/2}$, and when the diffusion coefficient $D_{\rm E}$ is in the energy-dependent regime ($E\ge1$~GeV), 
the escape length is given by $l_{\rm esc}\sim7.1\times10^{-4}(\tau_{\rm esc}/{\rm yr})^{1/2}(\nu_{\rm crit}/{\rm GHz})^{1/2}(B/{\rm \mu G})^{-1/2}$~kpc. During the above derived 
cooling time ($ \tau_{\rm esc}=\tau_{\rm cool}$) the electrons can travel only $l_{\rm esc}\sim  22-135$~pc. Hence, we conclude that if the FIR/star-forming sizes of our SMGs are as compact as those from Simpson et al. (2015a; $2.4\pm0.2$~kpc in median FWHM) and Ikarashi et al. (2015; $\sim1.6$~kpc in median FWHM), it seems unlikely that CR electrons would have had time to propagate from their sites of origin to the large distances where we observe the 3~GHz emission [the median major axis FWHM size being $ 4.2\pm0.9$~kpc]. In the above analysis we did not add the effect of 
the IC scattering off the CMB radiation, although at high redshift it can shorten the electron lifetime and diffusion length scale even more. However, 
apart from AzTEC3, our Fig.~\ref{figure:corr} does not show evidence of smaller radio sizes at higher redshifts as expected if the electrons have less time to 
travel away from their site of origin. This is possibly a manifestation of the fact that in starburst galaxies, at whatever redshift they might be, the local stellar radiation field is intense 
(cf.~the above estimate), and hence the CR electrons can suffer from strong IC losses from stellar light besides/instead from the CMB 
(e.g. \cite{lisenfeld1996}; \cite{lacki2010}). Moreover, we have ignored the fact that if the galaxy is associated with a galactic-scale wind, the CR particles in the 
wind adiabatically lose momentum and energy on the course of the expansion of the wind (e.g. \cite{volk1996}). The effect of losses due to electron advection would 
further shorten the diffusion length-scale.

\end{document}